\documentclass[useAMS,usenatbib]{mn2e}
\usepackage{graphicx}

\newcommand{\zsun}{Z$_{\odot}$}
\newcommand{\msun}{M$_{\odot}$}

\newcommand{\cii}{C~{\sc{ii}}}
\newcommand{\ciii}{C~{\sc{iii}}}
\newcommand{\civ}{C~{\sc{iv}}}
\newcommand{\siv}{S~{\sc{iv}}}

\newcommand{\siiv}{Si~{\sc{iv}}}
\newcommand{\pv}{P~{\sc{v}}}

\newcommand{\ovi}{O~{\sc{vi}}}

\newcommand{\niii}{N~{\sc{iii}}}
\newcommand{\hii}{H~{\sc{ii}}}
\newcommand{\heii}{He~{\sc{ii}}}
\newcommand{\ergs}{ergs s$^{-1}$ cm$^{-2}$ \AA$^{-1}$}

\title[Young Stellar Populations Observed with FUSE]{Physical 
       Properties of Young Stellar Populations in 24 
       Starburst Galaxies Observed with FUSE\thanks{Based 
         on observations made with the NASA-CNES-CSA Far
       	 Ultraviolet Spectroscopic Explorer. FUSE is operated for NASA by 
	 the Johns Hopkins University under NASA contract NAS5-32985.}}
\author[Pellerin \& Robert]{Anne Pellerin$^1$\thanks{E-mail: pellerin@stsci.edu} and Carmelle Robert$^2$\thanks{E-mail: carobert@phy.ulaval.ca} \\
$^1$Space Telescope Science Institute, 3700 San Martin Drive, Baltimore, MD 21218, USA. \\
$^2$D\'epartement de physique, de g\'enie physique et
d'optique, and Observatoire du mont M\'egantic, Universit\'e Laval, \\
Qu\'ebec, QC, CANADA, G1K 7P4}

\begin{document}

\date{Accepted 2007 July 19. Received 2007 July 18; in original form 2007 June 15}


\maketitle

\label{firstpage}

\begin{abstract}
We presents the main physical properties of very young stellar 
populations seen with FUSE in 24 individual starbursts. These 
characteristics have been obtained using the evolutionary spectral 
synthesis technique in the far-ultraviolet range with the {\tt LavalSB} 
code. For each starburst, quantitative values for age, metallicity, initial 
mass function slope, stellar mass, and internal extinction have been 
obtained and discussed in details. Limits of the code have been tested. 
One main conclusion is that most starbursts (and probably all of them) 
cannot be represented by any continuous star formation burst in the 
far-ultraviolet. Also, quantitative values of various optical diagnostics related 
to these stellar populations have been predicted. Underlying stellar populations, 
dominated by B-type stars, have been detected in NGC\,1140, NGC\,4449, 
and possibly NGC\,3991. We characterized the young stellar populations of 
less than 5\,Myr in Seyfert\,2 nuclei. 
\end{abstract}

\begin{keywords}
galaxies: evolution -- galaxies: stellar content -- galaxies: starburst -- galaxies: Seyfert -- line: profiles -- ultraviolet: galaxies.
\end{keywords}

\section{FUV Spectral Synthesis}

Since the introduction of the evolutionary population synthesis technique by 
\citet{tinsley68}, many codes have been developed to work with data at
various wavelength ranges, especially during the last decade 
\citep[e.g.][]{rob93,wor94,lei95,vaz99,lei99,molla00,rob03,bru03}.
These developments are mainly due to faster computers, advances in 
stellar evolution theories and also to progresses in space astronomy. 
Spectral synthesis of young stellar populations, as observed in luminous
far-infrared galaxies, nuclear starbursts, giant {\hii} regions, and also 
some Seyfert nuclei \citep{hec97a, gonz98a, gonz01}, has been particularly
successful in the ultraviolet from 1200 to 3000\AA\ thanks to several
space missions such as the {\it {Hubble Space Telescope}} (HST), the 
{\it {International Ultraviolet Explorer}} (IUE), and the {\it {Hopkins 
Ultraviolet Telescope}} (HUT).

Due to important instrumental constraints for the observation of the 
far-ultraviolet range (FUV; 900\AA\ $<\lambda<$ 1200\AA), very few 
synthesis codes have been developed in this regime. Until recently, only 
the {\tt Starburst99} code \citep[Gonz\'alez Delgado, Leitherer \& Heckman 
1997a;][]{lei99} was including a FUV spectral library based on HUT and 
{\it {Copernicus}} data. The lack of spectral resolution has restricted 
the synthesis to {\ovi~$\lambda\lambda$1032, 1038}, a doublet blended 
to Ly$\beta$ and {\cii} interstellar features. Furthermore, the low 
signal-to-noise ratio (S/N) of extragalactic data below 1200\AA\ has 
considerably limited the application of the technique in the FUV to a 
very few nearby objects (e.g. Leitherer, Calzetti \& Heckman 2002a). 

With the launch of the {\it {Far Ultraviolet Spectroscopic Explorer}} in 
1999 \citep[FUSE;][]{moos00,sah00}, the FUV regime can be fully
explored. Numerous O- and B-type stars have been observed and 
studied at those wavelengths \citep[e.g.][]{cro02,wal02,wil04}.
Comprehensive trends in the stellar line profiles of these stars have been 
found \citep{pel02}. \citet{rob03} have created new FUV spectral libraries 
of 228 OB~stars observed with FUSE for the {\tt LavalSB} and {\tt Starburst99} 
codes. Taking advantage of the impressive spectral resolution and sensitivity 
of FUSE, the authors have shown that, in addition to the {\ovi} structure 
studied by \citet{gonz97a}, other stellar indicators can be used for synthesis 
of very young populations in the FUV. The {\ciii~$\lambda$1176} line and the 
{\pv~$\lambda\lambda$1118, 1128} doublet have been revealed, in fact, more 
powerful indicators than the {\ovi}$+$Ly$\beta$$+${\cii} feature.

The new FUV library is a great opportunity to apply the spectral synthesis 
technique to various starbursts observed with FUSE. There are several 
advantages of the FUSE library on previous ones built in the same wavelength
range or based on HST/UV and Copernicus data. First, the spectral resolution is 
exceptional and leads to more accurate diagnostics based on the detailed 
stellar line profiles. Second, ions of {\ciii~$\lambda$1176} and 
{\pv~$\lambda\lambda$1118, 1128} do not show saturated profiles since 
{\ciii} is produced through excited atomic transitions and the {\pv} doublet
is of low cosmic abundance. This property allows a more accurate line profile 
fitting, especially at high metallicity, as it will be discussed in this work. Third, 
the flux calibration of FUSE data is more accurate than Copernicus data, which 
has some impact on the continuum normalization quality of the FUV stellar 
library. Fourth, the only stars contributing to $\lambda<$\,1200\AA\ are those 
hotter than 9200\,K ($\sim$A0-A2 type). Consequently, 
the FUV is not contaminated by the flux from older underlying generation 
of stars. This will allow us to determine physical properties strictly related 
to young stellar populations. 

The aim of this paper is to present FUV spectral synthesis on a relatively
large sample of nearby starburst galaxies. For each FUV observation, we 
will obtain precise quantitative values of several fundamental physical 
parameters on the young stellar populations (i.e. age, metallicity, initial 
mass function, mode of star formation, stellar mass, and internal extinction). 
We first present in the next section, the extragalactic FUSE spectra and 
summarize the data processing. In \S\ref{lavalsb} we briefly describe the 
{\tt LavalSB} code and the FUV library. Results from the synthesis modeling 
for each individual galaxy are  presented in \S4. We conclude in 
section~\ref{conclusion}.

\section[]{Observations and Data Processing}

The FUV spectra of the galaxies studied here have been taken from the FUSE 
archives through the {\it {Multimission Archive at Space Telescope}} 
(MAST\footnote[1]{http://archive.stsci.edu/fuse/}). One exception is 
NGC\,1667, for which A. Pellerin was the principal investigator. All FUSE 
spectra were collected with the 30\arcsec$\times$30\arcsec\ 
aperture (LWRS), except for NGC\,5253 (MDRS), between 1999 and 2002. 
Table~\ref{echantillon} summarizes the information relative to the galaxies and
their observations. The six columns of the table give the galaxy's name, the 
morphological type and activity, the central coordinates for the 
exposures, the Galactic extinction from NED\footnote[2]{{\it {NASA/IPAC 
Extragalactic Database}}; NED}, the distance, the FUSE project data set 
identification, and the total exposure time. The FUSE data have been 
processed using the {\tt calfuse} pipeline v2.2.2, the available version 
at that time. This version corrects for Doppler shift induced by the 
heliocentric motion of Earth, event bursts, the walk problem, 
gratings thermal shifts, bad pixels, background noise, distortions, and 
astigmatism. More information relative to {\tt calfuse} is available on the 
FUSE website\footnote[3]{http://fuse.pha.jhu.edu/analysis/calfuse.html}.
Flux calibration is accurate to better than 10\% in the LiF1A 
segment (between 987.1 and 1082.2\AA) and the dispersion is 
about 15-20\,km s$^{-1}$ pixel$^{-1}$.

Exposures obtained with the different detector segments have been 
combined in three steps. First, all exposures from the same segment have 
been added proportionally to their exposure time. Second, segments that 
cover the same wavelength range have been combined taking into account 
a statistical weight based on their signal-to-noise ratios (S/N). Note that a 
lower weight was given to the LiF1B segment (1094.3-1187.7\AA) because 
of the flux default around 1145\AA. This segment was not rejected because 
it covers the longer wavelength range where the stellar indicator 
{\ciii~$\lambda$1176} can be located. In a third step, each wavelength range 
has been co-added to obtain a single spectrum.
 
A correction for astrophysical redshifts was also applied following the 
heliocentric radial velocities given by NED. Finally, the spectra were
smoothed with a 20 pixels box using the task
{\it {boxcar}} in IRAF\footnote[4]{{\it {Image Reduction and Analysis Facility}}; 
http://iraf.noao.edu/}. This last operation increases the S/N without 
affecting the stellar line resolution needed for the spectral synthesis.
The FUSE spectra presented in this work are electronically available on the 
MAST archives\footnote[5]{http://archive.stsci.edu/prepds/fuse\_galaxies/\,.}.

\section{Evolutionary Spectral Synthesis with LavalSB}
\label{lavalsb}

{\tt LavalSB} is an evolutionary synthesis code for young stellar populations 
developed in parallel with {\tt Starburst99} \citep{lei99}. A description of 
{\tt LavalSB} in relation with its application to the synthesis in the FUV range 
can be found in \citet{rob03}.
In summary, {\tt LavalSB} uses the evolutionary tracks of the Geneva group 
\citep{schal92,schae93a,schae93b,cha93,mey94}. Initially the stellar 
population obeys a mass distribution based on a chosen initial star formation
scenario, i.e. an initial mass function (IMF) and mode of star formation
(instantaneous or continuous). Individual stellar parameters are considered
to assign the corresponding normalized empirical spectrum from the FUV 
library based on relations from \citet{schm82}. The normalized library 
spectra are flux calibrated using stellar atmosphere models of \citet{kur92} 
and of \citet{sch92} for stars with extended envelop. Prior to their use, 
the \citet{kur92} spectra have been fitted using a Legendre function in 
order to remove their low resolution spectral features. The code also 
includes evolutionary tracks for massive close binary stars, which was not 
used for the current work since binary stars do not have 
significant impact on UV line profiles \citep{dionne99,dionne05}.

The FUSE stellar library covers from 1003.1 to 1182.678\AA\ with a 
dispersion of 0.127\AA. For {\tt LavalSB}, this library is composed of 
155 Galactic stars from O3 to B3 spectral types, as well as 41 and 
32 stars from the Large and Small Magellanic Clouds (LMC and SMC), 
respectively, which includes spectral types between O3 and B0. 
Wolf-Rayet (WR) stars are also included.
As pointed out by \citet{rob03}, the lack of stars cooler than B0 at 
sub-solar metallicities produces a significant dilution of stellar line 
profiles for ages older than 7.0\,Myr. The library metallicity is matched
with the corresponding evolutionary tracks in {\tt LavalSB}. For the LMC 
stars, we assume a metallicity of 0.4\,\zsun (i.e. 12+log[O/H]=8.3, where 
the metallicity for solar environment objects corresponds to 
12$+$log[O/H]=8.7). For SMC stars, we assume a metallicity around 
0.2\,\zsun (i.e. 12+log[O/H]=8.0). Evolutionary models at 2\,{\zsun} are 
also produced with {\tt LavalSB} but using the spectral library at solar 
metallicity. The reason is that not enough massive stars can be 
observed with FUSE with such a high metallicity. Below 1200\AA,
the strength of stellar wind lines is significantly changing at high 
metallicity, unlike for $\lambda>$ 1200\AA\ where the profiles are 
quickly saturated. The consequences of using the solar metallicity 
library for objects with higher metallicity for FUV synthesis is not 
negligible and will be discussed in \S4.

The work from \citet{rob03} has revealed that synthetic FUV spectra of young 
stellar populations contain important stellar indicators. The most useful 
ones are the {\ciii} multi-line centered at 1176\AA, and the {\pv} doublet 
at 1118 and 1128\AA. The shape of these lines shows strong variations 
with age and metallicity. Significant, but more subtle, changes also appear 
with different IMF parameters. For young stellar population of less than 
$\sim$10\,Myr, at least, there is no age-metallicity degeneracy observed 
in the FUV; the age influence the global P\,Cygni profile while the 
metallicity mostly affects the line depth. At shorter wavelengths, the 
{\ovi~$\lambda\lambda$1032, 1038} and the {\siv~$\lambda\lambda$1063, 
1073} line profiles show variations with the age and metallicity, and
possibly with the IMF. However, these diagnostic lines are contaminated by 
interstellar features of H$_2$ and other atomic transitions and are 
consequently more difficult to use for the spectral synthesis. Nevertheless, 
the {\ciii~$\lambda$1176} and {\pv~$\lambda\lambda$1118, 1128} lines 
together are powerful indicators to find physical parameters 
of young stellar populations, as it will be shown in \S4.

\section{Properties of Young Stellar Populations in the FUV} 

In this section, we present detailed results obtained from spectral 
synthesis for each individual starburst galaxy listed in Table~\ref{echantillon}.
This work is part of \citet[][Ph.D. Thesis]{pel04} where more details and figures
can be found . The FUV synthesis is a three step process.
First, for each normalized spectrum, the observed {\pv} and {\ciii} line 
profiles are compared to various models produced with {\tt LavalSB}. 
The models cover stellar populations with different IMF slopes (using mass
cutoffs of 1 and 100\,\msun), and either instantaneous (where all stars are
formed at the initial time) and continuous (where new stars are added to the
population at each time step) modes of star formation. The comparison 
between the models and the normalized FUSE spectra is done both by eye and 
using a $\chi^2$ method. In the case of noisy spectra, an eye judgment only is
often unavoidable. This first step gives quantitative values, along with 
their uncertainties, for the age, the metallicity, and the IMF slope. 

In a second step, the continuum level and its slope are studied. The spectra are
first corrected for the Galactic reddening using the \citet{card89} law. The 
internal extinction E(B$-$V)$_i$ for each starburst galaxy is evaluated from
the FUV continuum slope using the theoretical extinction law from 
\citet{witt00} for a clumpy dust shell distribution with an optical depth 
of 1.5 in the V band. According to \citet{buat02}, this law is in agreement 
with the observed FUV spectral energy distribution of starbursts. More 
precisely, the internal extinction is obtained by comparing the observed 
continuum slope $\beta_{obs}$ (where F$_{\lambda} \propto \lambda^{\beta}$) 
with the synthetic slope $\beta_{th}$ for the best model found from line profiles. 
The FUV slopes are measured through 5 small bandwidths that avoids stellar 
and major interstellar features (see Tab.~\ref{slopebands}). Also, we restricted 
these bands to the longer wavelengths of the FUSE spectrum since the 
continuum below 1200\AA\ is not always well represented by a power-law, 
especially at smaller wavelengths. We used the IRAF routine {\it {curfit}}
with different values for the low and high pixel rejection levels to obtain an average
value of the continuum slopes. Although this method gives uncertainties for the slopes, 
and therefore for the extinction, we also preformed a direct eye comparison of the galaxy
extinction corrected spectrum with the synthetic spectrum. For galaxies with low     
signal-to-noise ratio spectra and those with a higher redshift (i.e. with limited
spectral coverage at longer wavelength) larger uncertainty values for the
extinction have been adopted, revealing still reasonable  superposition of the
spectra. Once a correction for the internal extinction is applied, the stellar mass M$_{\star}$
involved in the observed burst is calculated. To do so, one must take into account 
the galaxy distance and then calculate the multiplying factor needed to 
superimpose the best-fitting synthetic spectrum obtained from the line 
profile to the flux level of the corrected spectrum. {\bf Since the stellar mass is 
calculated after correction for internal extinction, the stellar mass value is highly 
dependent on the extinction evaluation, which depends strongly on the 
quality of the spectra and its spectral interval covage. As it will be 
discussed later in the text, the stellar mass values derived from FUV luminosities 
are usually in good agreement with other works, but can lead to large 
uncertainties (and nonsense values) when the spectra quality is poorer.
For example, an error of 0.1 in E(B-V) will lead to a variation of 
Log(M$_{\star}$) of $\pm$1.}

The third step uses {\tt LavalSB} to compute and predict other galaxy 
observables based on the best-fitting model. We compute the flux and 
equivalent width (EW) of the nebular line H$\alpha$, the number of O, B, 
and WR stars, the flux of the WR bump around 4686\AA, and the 
continuum flux level at 5500\AA. These predictions are then compared 
to observations found in the literature. Taking into account the different
apertures used, such comparisons often offer a good test for the FUV 
synthesis.

In the following sub-sections, we first discuss the case of well-known
starburst galaxies for which we can compare the FUV synthesis results
with analysis at other wavelengths from the literature. In \S\ref{metal},  
we discuss the impact of discrete metallicity values available in the code 
on integrated stellar populations with an intermediate metallicity. We then 
perform, in \S\ref{lowz}, the FUV synthesis of relatively low metallicity 
objects and discuss the limitation of the FUV library. In \S\ref{bstarimf} we 
present the case of stellar populations for which a non-standard IMF slope 
has been obtained. The FUV synthesis of low signal-to-noise ratio spectra 
is described in \S\ref{othersb}. Finally, in \S\ref{synsyft} and \S\ref{syft1}, we 
summarize the synthesis result for Seyfert galaxies, some of which showing 
a FUV spectrum dominated by a young stellar population and others
dominated by their non-thermal emission.

\subsection{FUV Synthesis of Well-Known Galaxies: }
\label{prototypes}

{\bf NGC\,7714} is a spiral galaxy often described as the prototype for 
nuclear starbursts  \citep{weed81,gonz99}. The FUSE aperture is 
centered on the nuclear region, covering a field of 
5.7$\times$5.7\,kpc$^2$ (at a distance of 37.5\,Mpc; see 
Table~\ref{echantillon}). Figure~\ref{figproto} shows the good 
quality spectrum of NGC\,7714 obtained by FUSE (with 
S/N$\simeq$15 at 1150\AA).  
The diagnostic line of {\ciii} displays a strong photospheric profile 
as well as a weak but broad absorption in the blue wing which is an 
indication of stellar wind from O-type stars. The {\pv} doublet shows 
deep and well-defined absorptions. The best model for the {\pv} and 
{\ciii} line profiles, as shown in Figure~\ref{figproto}, is obtained for an 
instantaneous burst of 4.5$\pm$0.3\,Myr at solar metallicity, and using 
a standard IMF ($\alpha$=2.35, from 1 to 100\,{\msun}). 
Models with an age of 4.0 and 5.0\,Myr are significantly less satisfying
to reproduce the line profiles. Models with another metallicity do not 
give the right depth for the lines. Note that a different IMF slope between 
$\alpha$=2.2 and 3.3, can also fit the data, but because
of the spectrum noise level, we simply adopt here a standard IMF for our study.
A continuous burst of $\sim$10-20\,Myr can also reproduce the line profile, 
but one must be careful in this specific case before concluding. 
Indeed there is a degeneracy between some continuous burst models 
and an instantaneous burst of $\sim$4.5\,Myr. This effect is not
seen for an instantaneous burst at a different age. For other galaxies 
of our sample having a different age and enough signal in their FUSE 
spectrograms, the case of a continuous burst is 
generally discarded. For this reason, we favor the instantaneous 
model for NGC\,7714. The best model parameters retained from 
the FUV synthesis are reported in Table~\ref{allsyn}.

An age of 4.5\,Myr is in agreement with previous studies using other
wavelength bands \citep{gonz99,gar97,gonz95}. An ambiguous point in 
the literature is the metallicity of NGC\,7714. \citet{gonz95} measured 
12+log[O/H]=8.5 (i.e. 0.6\,\zsun) for the nucleus, while \citet{hec98} 
compiled from other papers an oxygen abundance higher than solar.
More recently, \citet{fer04} obtained 12+log[O/H]=7.9 (i.e. 0.2\,\zsun). 
The limited precision in metallicity of {\tt LavalSB} do not allow us 
to distinguish between the values of  \citet{gonz95} and \citet{hec98},
but certainly the FUSE synthesis do not agree with the metallicity 
of \citet{fer04}.

Based on the best-fitting model for the line profiles, the FUV continuum 
slope should be $\beta_{th}$=$-$1.8$\pm$0.2. A comparison with the 
FUSE spectra allow us to estimate an internal extinction 
E(B$-$V)$_{i}$=0.1$\pm$0.1, if we consider a Galactic contribution 
E(B$-$V)$_{Gal}$=0.08, as proposed 
by \citet{gonz99}. The UV continuum slope and the 
flux level at 1150\AA, after correction for the Galactic extinction, are given 
in Table~\ref{allsyn}. This is consistent with result of 0.03 obtained from 
HST/GHRS spectra by \citet{gonz99}. One important difference with this 
study is the aperture size; the HST/GHRS aperture is 
1.74\arcsec$\times$1.74\arcsec\, which is much smaller than the 
FUSE aperture.

When correcting the FUSE spectrum with the internal extinction of
E(B$-$V)$_i$=0.1, we estimate that the total stellar mass responsible for 
the observed FUV flux is of the order of 10$^8$\,{\msun}. The mass 
uncertainty is mainly related to the extinction uncertainties. These results 
are consistent with 
a mass of 6.8$\times$10$^7$\,{\msun}, derived by \citet{gonz95} from the 
H$\alpha$ intensity (1.2\arcsec\ slit), and 1.1$\times$10$^8$\,{\msun}, from 
\citet{gonz99} based on the IUE (10\arcsec$\times$20\arcsec) flux observed 
at 1500\AA. The mass differences can be easily explained by the various 
internal extinctions and aperture sizes used. 

Using the starburst physical parameters found from the FUV synthesis, 
it is possible to predict several features in the visible range. Table~\ref{predictions} 
gives, for example, the number of O stars 
and the WR/O ratio obtained from the best model, along with the predicted 
flux and EW for the H$\alpha$ line, the equivalent width of the WR bump at 
4686\AA, and the continuum flux level at 5500\AA. In the central 330\,pc 
(1.74\arcsec) of NGC\,7714, \citet{gonz99} derived about 16\,600 O~stars. 
The best model for the FUSE data predicts 12 times more O~stars within the 
central 30\arcsec$\times$30\arcsec. The large difference in aperture sizes do 
not  allow us to compare these numbers. The same authors derived a WR/O 
ratio of 0.12, relatively close to the ratio of 0.19 derived with FUSE. Note that 
this last ratio is very sensitive to the age, i.e. at 5.0\,Myr LavalSB gives 0.12. 
\citet{gonz95} also detected a WR bump in a 1.2\arcsec\ slit with 
EW(4686)=1\AA, a much lower value than 44\AA\ as predicted for the FUSE 
aperture. The H$\alpha$ flux and equivalent width of NGC\,7714 have been 
measured by several authors. If we limit our comparison with the 
1.9$\times$10$^{-12}$\,{\ergs} H$\alpha$ flux value of Storchi-Bergmann, 
Kinney \& Challis (1995) in a large 10\arcsec$\times$20\arcsec\ aperture, 
the value predicted from the 
FUV is very consistent. \citet{stor95} also measured a continuum flux
F(5550)=1.4$\times$10$^{-14}$\,{\ergs}, which corresponds well again to 
the FUSE prediction of 10$^{-14}$\,{\ergs}.


{\bf IRAS\,08339+6517} is a spiral galaxy with a strong and compact 
nuclear starburst \citep{mar88} which was probably triggered by an interaction
with 2MASX\,J08380769$+$6508579 \citep{can04}. Because of its relatively large distance 
(78\,Mpc), the {\ciii} structure is redshifted outside the FUSE spectral 
range, as seen in Figure~\ref{figproto}. The high S/N of $\sim$15 allows 
a good synthesis of the {\pv} doublet. From the {\pv} line profiles, we find 
an age of 7.0$\pm$0.3\,Myr for an instantaneous burst at {\zsun}. A 
standard IMF is adopted, but a good fit is also obtained for an IMF slope 
of 2.8. A continuous burst cannot reproduce the line profiles since this 
scenario would always show  P\,Cygni profiles in {\pv}, 
which are not observed in IRAS\,08339+6517. 
The age and metallicity from the FUV synthesis are in really good 
agreement with the previous work of \citet{mar88} in the visible, 
\citet{gonz98a} and \citet{lei02} using UV spectra from HUT and HST, 
and \citet{pel99} with visible and near-infrared spectra. 

Using our best-fitting model for the FUV line profiles, we calculate an 
internal extinction E(B$-$V)$_i$=0.30$\pm$0.15, with E(B$-$V)$_{Gal}$=0.092 
(NED). The large uncertainty comes from the fact that the rest-frame FUV spectrum 
is incomplete at longer wavelength due to the relatively high redshift of 
IRAS\,08339$+$6517. A stellar mass of 10$^{8}$ to 10$^{12}$\,{\msun} is then 
derived for the young burst. {\bf We can see here the effect of the shorter spectral 
coverage on the mass estimate, which lead to large uncertainties. A mass of 
10$^{11}$-10$^{12}$\,{\msun} is similar to the mass of an entire spiral 
galaxy like the Milky Way. Such a mass is clearly unreasonnable in the 
case of IRAS\,08339$+$6517 since the FUSE aperture does not even cover
the full galaxy. The stellar mass values are obviously
less reliable when the uncertainties on extinction increase.}
\citet{gonz98a} and \citet{lei02} obtained E(B$-$V)$_i$=0.17 from the HUT 
data (using E(B$-$V)$_{Gal}$=0.08, which is similar to the value we used).
Their internal extinction is fairly consistent within the lower limit of the value 
obtained from the FUV. {\bf For this reason, we suspect that the stellar mass for 
the nuclear starburst in IRAS\,08339$+$6517 is probably closer to 10$^{8}$\,\msun.}

Predicted values for visible features for IRAS\,08339$+$6517 are reported in 
Table~\ref{predictions}. \citet[with a 22\arcsec\ diameter aperture]{mar88}, 
\citet[10\arcsec$\times$20\arcsec\ aperture]{gonz98a}, and 
\citet[4\arcsec\ slit]{pel99} observed an H$\alpha$ flux of 0.7, 1.3, and 
1.5$\times$10$^{-12}$\,{\ergs}, respectively. These are low compared
to the predicted value of 10$^{-10}$\,{\ergs}, but in agreement considering 
the large uncertainties. The flux level at 5500\AA\ observed by \citet{pel99} is 
1.8$\times$10$^{-14}$ {\ergs}, consistent within the lower limit predicted from
FUSE data. It may also indicate that FUSE, with its larger aperture, is seeing 
more than just a central compact starburst.


{\bf M\,83 (NGC\,5236)} is a well studied spiral galaxy at a distance of 
3.8\,Mpc. The FUSE aperture covered the circumnuclear ring of star 
formation. In Figure~\ref{figproto}, the observed {\ciii} line clearly displays a
broad absorption in its blue wing, which is associated with 
evolved O~stars. The important depth of {\ciii} and {\pv} suggests 
metal rich stars. While {\tt LavalSB} has evolutionary tracks at 
2\,{\zsun}, it does not have a stellar library at this metallicity (see 
\S\ref{lavalsb}). Nevertheless, a relatively good fit at 2\,{\zsun} is found, 
which can satisfy the width and the emission part of the
P\,Cygni profiles for an age of 3.5$\pm$0.5\,Myr.
As shown in Figure~\ref{figproto}, the line depth cannot be reproduced
because of the lack of metal rich stars in the FUV library.
However, we can assert that the metallicity is significantly higher than solar.
A standard IMF was simply adopted here because of the impossibility
to properly consider the line depth.
A continuous burst model cannot be adjusted to the FUSE spectrum 
of M\,83 mainly because the synthetic {\pv} doublet shows P\,Cygni 
profiles which are too week compared to the observed lines.

The age obtained from the FUV synthesis is in good agreement with
other observations \citep{elm98,pux97,har01,lei02}.
In their detailed work, \citet{bre02} studied individual knots of star formation
around the galaxy center.  The brightest and youngest knot, named~A,
displays a 4\,Myr old population with evidence for WR
stars. It seems therefore that knot~A contributes significantly to the FUSE
spectrum. The high metallicity obtained from the FUV is 
consistent with the work of Zaritsky, Kennicutt \&
Huchra (1994) and \citet{bre02}  who found an
oxygen abundance of 12$+$log[O/H]=9.2 (i.e. 3.2\,{\zsun}). 

From the FUV continuum slope, we measure an intrinsic extinction
E(B$-$V)$_i$=0.08$\pm$0.08, using E(B$-$V)$_{Gal}$=0.066 (NED).
From the Balmer decrement, Storchi-Bergmann, Calzetti \& Kinney (1994) 
and \citet{cal95} measured a total extinction of 0.29, consistent with 
the FUV extinction. Although the aperture sizes are quite different, 
it is normal to find a UV extinction that is lower by 50\%
compared to the extinction from the Balmer decrement \citep{cal95}.
After corrections for the reddening, the FUV flux gives a stellar mass
between 2.5$\times$10$^5$ and 10$^7$\,{\msun}. The work of \citet{bre02} 
revealed a stellar mass of $\sim$10$^5$\,{\msun} for knot~A only, which 
strongly suggests again that it is a major contributor to the observed 
FUSE spectrum.

\citet{bre02} also studied the WR content in M\,83. For knot~A, they 
measured EW(4686)=4.6\AA\ from which they estimated the presence of 
31 WR stars. {\tt LavalSB} predicts EW(4686)=18\AA\ 
and about 2000 WR stars (see Tab.~\ref{predictions}). This 
discrepancy can easily be explained by the fact that the FUSE aperture
includes other young knots and also because the WR bump at 4686\AA\ 
can underestimate the number of WR stars 
due to  the contribution of other stellar populations to the visible
continuum \citep{chan04}. 
\citet{cro04} identified 1030 WR stars for the whole galaxy, and this number
might reach 1500, according to their prediction. FUV synthesis results are 
consistent with their observations, considering the large uncertainty of the
predicted values and the very different apertures.

Bresolin (2003; private communication) measured 
F(H$\alpha$)=1.34$\times$10$^{-12}$\,{\ergs}
and F(5500)=6.94$\times$10$^{-15}$\,{\ergs} for knot A alone, yet showing 
the importance of this knot to the luminosity in the nuclear region
based on the FUV predictions (see Tab.~\ref{predictions}).
\citet{stor95}, using a 10\arcsec$\times$20\arcsec\ aperture,
measured an H$\alpha$ flux which is 10 time smaller than
the one predicted by {\tt LavalSB}.
On the contrary, \citet{stor95} measured a continuum flux at 5500\AA\  which
is twice as strong as the one predicted from the FUV.
This suggests that there is an important contribution to the 
visible continuum flux from an older stellar populations (too cold to be seen
in the FUV) which is diluting the H$\alpha$ line flux.


{\bf NGC\,3690 (Mrk\,171)} is a member of the highly perturbed interacting
system Arp\,299 \citep{hib99}. NGC\,3690 and its companion, IC\,694, are both
displaying intense starbursts. The FUSE aperture includes most the emission 
from NGC\,3690 and excludes IC\,694. As shown in Figure~\ref{figproto}, the 
FUV spectrum displays an incomplete {\ciii} line. Nevertheless, the blue wing 
of the {\ciii} line is still very useful for a FUV synthesis. Models for a instantaneous 
burst at solar metallicity are best to reproduce the {\ciii} and {\pv} line depths.
While the {\ciii} line favors models with an age between 5.5 and 6.5\,Myr, the 
{\pv} feature is better adjusted by a burst of 6.5-7.0\,Myr. The best-fitting model 
of 6.5$\pm$0.5\,Myr also indicate a standard IMF. For example, a steeper IMF 
slope is rejected because it cannot reproduce the weak blue wing observed for 
the {\ciii} line.

From visible and IR data, ages between 3.5 and 8\,Myr have been found
for individual bright knots in NGC\,3690
\citep[e.g. Gehrz, Sramek \& Weedman 1983;][]{bon99,saty99,alon00}.
\citet{schae99} detected WR star signatures in a region inside
the FUSE aperture which also indicates a young age.
\citet{rob99} performed the UV synthesis of a HST/FOS spectrum
obtained with a 1\arcsec\ aperture centered on knot B2 (included in the
FUSE aperture) and found an age of 6.5\,Myr.
A metallicity of 1.26\,{\zsun} was measured by \citet{hec98}
which is also consistent with the FUV result.

Using the best model above, we measured an internal extinction
E(B$-$V)$_i$=0.15$\pm$0.05, when adopting E(B$-$V)$_{Gal}$=0.017 (NED).
After correction for the extinction, the FUV flux level gives a stellar
mass of 10$^8$\,{\msun} for the young
burst. \citet{alon00} estimated a mass of 3$\times$10$^7$\,{\msun} for
the small knot~B2 alone. Table~\ref{predictions} reports values of the visible 
flux and line equivalent widths predicted from the FUV synthesis.
Although generally consistent with the literature, these numbers must be 
compared with caution mainly because they have been collected using a 
different aperture size. For example, \citet{geh83} used a small 4\arcsec\ slit 
and measured a value for the H$\alpha$ flux which is 5 times weaker than 
the one predicted. \citet{ken92} found a closer number with a circular aperture 
of 45\arcsec\ diameter. \citet{alon00} and \citet{fri87} gave H$\alpha$ equivalent 
width measurements between 55 and 600\AA\  for various knots, which brackets 
the FUSE prediction of 140$^{+180}_{-120}$\AA.


{\bf NGC\,3310} has a very blue compact nucleus and a 
circumnuclear starburst. The FUSE aperture contains both structures 
and allows us to study most of the galaxy
star formation activity since it is estimated to take place within the central 
20\arcsec\ \citep{smi96}. Figure~\ref{figproto} shows a high quality
FUV spectrum for NGC\,3310 with a S/N$\simeq$20. Note that the stellar 
{\pv~$\lambda$1118} line is contaminated by Galactic interstellar lines of 
H$_2$. The {\pv~$\lambda$1128} component and the {\ciii} line do not show any 
extended blue absorption from stellar wind. From this, we conclude that no 
significant O~star population is present. We find that the 
best line profile adjustment is obtained with an instantaneous burst of 
18$\pm$2\,Myr at 2\,{\zsun} and a standard IMF. Another model
can reasonably fit the lines, which consists of a 
15$\pm$1\,Myr burst at {\zsun}, if a steeper IMF 
slope of 2.80 is used. It would probably be possible to separate the two 
cases if a proper stellar library at 2\,{\zsun} was available. An abundance of
12+log[O/H] = 8.8-9.0, corresponding to 1.3-2.0\,{\zsun} was found by 
\citet[Denicol\'o, Terlevich \& Terlevich 2002 and][]{hec98}. This favors 
the model at 2\,{\zsun}. Furthermore, no continuous 
burst model can reproduce the observations.

Only a few studies addressing the stellar content in the nucleus of 
NGC\,3310 are found in the literature. Most suggest an age between 6 
and 20\,Myr \citep{boi00,lei02,elm02,hug05}. However, \citet{pas93} found a 
younger age for a small region located about 10\arcsec\ (SW) from the 
nucleus, based on the observation of a WR bump at 4686\AA. However, 
the FUSE spectrum do not show evidence of a stellar population
younger than 10\,Myr probably because these young stars do 
not contribute significantly to the FUV flux within the large aperture.
These stars seem to be 
completely diluted by a more important B-type star population.

Considering the best-fitting model at 2\,{\zsun}, we estimate an internal 
extinction E(B$-$V)$_i$=0.05$\pm$0.05 (using E(B$-$V)$_{Gal}$=0.022)
and a stellar mass of 0.39-3.9$\times$10$^8$\,{\msun}. The stellar mass of 
2.2-8.4$\times$10$^8$\,{\msun} calculated by \citet{smi96} within a 
$\sim$25\arcsec\ inner region is in agreement with FUV result.
Table~\ref{predictions} reports the predicted fluxes and EW in the visible
for NGC\,3310. \citet{pas93} measured a continuum flux levels at 5510\AA\ 
for individual regions. If we add together their fluxes for the seven knots 
included within the FUSE aperture, we obtain a lower limit of 
F(5500)$>$1.5$\times$10$^{-14}$ {\ergs}, 
which is consistent with the synthetic flux between
2.5 to 25$\times$10$^{-14}$ {\ergs}.
However, the predicted flux for H$\alpha$ is significantly lower than 
the 1.4$\times$10$^{-12}$\,{\ergs} measured by \citet{pas93} 
for the same seven knots, suggesting a contribution to H$\alpha$ 
from other sources not seen in the FUV.

\subsection{Adjusting the Metallicity}
\label{metal}


{\bf NGC\,4214} is a relatively close (3.4\,Mpc) irregular galaxy with
many knots of intense star formation. Doing the FUV synthesis of its FUSE spectrum
(shown in Fig.~\ref{figmetal}) was not easy at first. Despite a high S/N of $\sim$19,
we were unable to obtain an adequate adjustment of the stellar lines using  
the models at the available metallicities. Mainly, models at 0.4\,{\zsun} produce 
stellar lines which are too shallow while lines from models at {\zsun} are too deep. 
Among these, the best model (but still not satisfying) was obtained for an 
instantaneous burst of 4.5-5.0\,Myr  at 0.4\,{\zsun} with $\alpha$(IMF)$\leq$2.35.

\citet{kob96} report an abundance 12+log[O/H] = 8.2 (i.e. 0.3\,{\zsun}) for 
NGC\,4214, which is consistent with a modeling at low metallicity. Furthermore, 
these authors observed an abundance gradient. Considering this spatial 
variation and the existence of multiple knots of star formation, we then performed 
a more sophisticated synthesis. We compared the observed FUV spectrum 
with the synthetic spectrum of a double burst model. This model was created by 
combining two synthetic spectra of two different bursts, one at a metallicity of 
0.4\,{\zsun}, and the other at {\zsun}. A standard IMF was assumed here.
The comparison of the observed spectrum with those obtained with the
double bursts model are shown in Figure~\ref{fig4214}. The {\pv} and 
{\ciii} line profiles are better reproduced if we consider an age of 5.5\,Myr 
and a similar flux proportion for the two bursts. Models of 5.0 and 6.0\,Myr 
have more difficulty to reproduce the blue wing of {\ciii}, but still give 
reasonable fits. Other flux proportions between the two bursts are less
good at reproducing the line profiles. In conclusion, the global metallicity 
for the stellar population seen with FUSE in NGC\,4214 seems to be slightly 
higher than the LMC metallicity. A detailed study of the stellar content of 
NGC\,4214 has been performed by \citet{lei96}, revealing around 200 
point sources, apparently individual hot stars of small clusters, 
distributed around a more important starburst. Their UV synthesis 
based on an  HST/FOS spectrum (1\arcsec\ circular aperture) on the central 
burst leads to an age of 4-5\,Myr, which is in good agreement with 
our results. 

The spectral energy distribution of NGC\,4214 shows some fluctuations.
This problem was also observed for NGC\,5253 (from the same observing 
program) and Mrk\,153. Compared to the synthetic energy distribution, the 
continuum level at $\lambda>$1100\AA\ is too high, and the level between 
1000 and 1100\AA\ is too low (not shown here). This behavior cannot be 
attributed to the stellar population itself or the ISM. It is possibly related to 
the alignment of the FUSE spectrograph optical paths. If this alignment is 
not perfect, especially for extended targets, the flux can be slightly different 
from one segment to another, as seen here. The consequence is to increase 
the uncertainties on the extinction and the stellar mass.

Despite these fluctuations, we estimate an internal extinction of 0.13$\pm$0.03 
using a E(B$-$V)$_{Gal}$=0.022 (NED), which is in agreement with 
the measurement of \citet{lei96}. This leads to a stellar mass of 
(2$\pm$2)$\times$10$^6$\,{\msun} for the young burst.
Predicted fluxes and EW for H$\alpha$ and continuum observables
(Tab.~\ref{predictions}) are in good agreement with the observations 
of \citet[][using a 40\arcsec\ diameter aperture]{mac00}. 
\citet{sar91} observed WR bumps in 4 bright knots and estimated that about 
233 WR star were present in those knots. This is only 25\% of the value
predicted by {\tt LavalSB} for a 5.5\,Myr population. The difference suggests
that an important number of WR stars resides in other clusters.


{\bf NGC\,5253} is experiencing a violent star formation event possibly 
due to its interaction with M\,83 \citep{cal99}. As shown in 
Figure~\ref{figmetal}, the {\ciii} and {\pv} lines in the FUSE 
spectrum of NGC\,5253 and NGC\,4214 are quite similar, showing the same
line depth and no P\,Cygni profile. The synthesis is again
indicating an age of $\sim$4.5\,Myr, with a metallicity around 0.4\,{\zsun},
but the result is not fully satisfying. The difficulty encountered here is identical to the 
one for NGC\,4214 and probably has the same origin since both objects have
a similar metallicity \citep{den02}. Therefore, we are considering synthetic spectra
which are composed of a blend of two models with metallicities of 0.4\,{\zsun}
and {\zsun}. Our best synthetic spectrum (see Fig.~\ref{fig5253}) indicates, 
assuming a standard IMF, a flux proportion of about 75\% for the 0.4\,{\zsun} 
population and about 25\% for the {\zsun} population, with an age of 
5.5$\pm$1.0\,Myr for both populations. In conclusion, we estimate 
that the global stellar population of NGC\,5253 is well represented with a
metallicity between 0.4 and 0.5\,{\zsun}. This idea of an intermediate metallicity 
was also raised by \citet{tre01} while analyzing visible data with {\tt Starburst99}.
This age found is consistent with previous works \citep{cal97,tre01}. 
\citet{lei02} estimate an average age of 20\,Myr with HUT spectra and a 
10\arcsec$\times$56\arcsec\ aperture, and they assert that 
O~stars, while present, are not dominant in this object, which is
different from the FUV point of view.

Using the Galactic extinction values of 0.056 (NED), we find an internal extinction
of 0.05$\pm$0.05. With this correction, we estimate a stellar mass of 0.8 to 
7.9$\times$10$^5$\,{\msun}.  The uncertainty for the stellar mass also takes into
account the problem seen for the flux distribution, as discussed for NGC\,4214.
A value of 6.6$\times$10$^4$\,{\msun} was obtained 
by \citet{tre01} using a long slit, while a mass of 10$^5$\,{\msun} 
was calculated by \citet{cal97} for the five brightest knots together. These values
are consistent with each other, considering differences in apertures and 
extinction values.

The continuum level at 5500\AA\ obtained by \citet{stor95}, with a 
10\arcsec$\times$20\arcsec\ aperture, is of 2.4$\times$10$^{-14}$ {\ergs}. 
This is consistent withh the predictions (see Tab.~\ref{predictions}).
The predicted H$\alpha$ flux is also in agreement with the value of 
7.7$\times$10$^{-12}$ from \citet{cal97} for the five brightest clusters 
included within the FUSE aperture. It is interesting to note that that
the brightest knot (\#5) alone produces an H$\alpha$ flux of
5.8$\times$10$^{-13}$\,{\ergs}. This knot has a 0.35 mag of internal
absorption between H$\alpha$ and H$\beta$, which certainly leads to a 
strong obscuration of this knot at UV wavelengths.
\citet{schae97} estimated the number of hot stars in NGC\,5253. They 
obtained 1730 and 840 O~stars in their region~A and B, respectively, and 
a WR/O ratio of 0.02 and 0.06 for the same two regions. The number of O-type
stars consistent with the predicted values of {\tt LavalSB}, while the number
of WR stars is very sensitive to the age of the best model. 

\subsection{Low Metallicity Starbursts and the Limit of the FUV Library}
\label{lowz}


{\bf Mrk\,153}, although less studied, is a well established
starburst galaxy.  Its FUV spectrum is of good quality,
with  S/N$\simeq$19 at 1150\AA, except in the
LiF1B segment around the {\ciii} line where it becomes more noisy
(see Fig.~\ref{figlowz}). Its {\pv} and {\ciii} lines are similar to those of
NGC\,4214 and NGC\,5253 (Fig.~\ref{figmetal}) with no sign of wind
profiles (i.e. extended blue wings). Note that the 
{\pv~$\lambda$1118} component is possibly contaminated by 
an airglow line. Shallow stellar lines are usually an indication of
a metal poor population. 
Based mainly on the {\pv~$\lambda$1128} feature, a good model
might be an instantaneous burst of 6.5$\pm$1.0\,Myr at 0.2\,{\zsun},
assuming a standard IMF. A model at 0.4\,{\zsun} is rejected as
it produces absorption lines in {\pv} that are too deep. However, as 
shown in Figure~\ref{figlowz}, the best model cannot reproduce the depth of the
{\ciii} feature. Although the S/N is lower in this wavelength range, we
believe it is an indication of an older age for the burst. Around this age,
synthesis work at low metallicity is limited by the fact that stars cooler
than B0 are missing in the spectral library (see \S\ref{lavalsb}).
Because of this limitation, we did not test other IMF slopes.
\citet{kun85} have found a metallicity of 0.1\,{\zsun}, i.e. 7.8 in 12+log[O/H].
Also, based on an IUE spectrum, \citet{bon99} conclude to
an age $<10$~Myr. These results are consistent with the FUV
synthesis.

Based on the FUV slope for a 6.5~Myr old stellar population, we calculate
an intrinsic extinction of 0.07$\pm$0.07, with E(B$-$V)$_{Gal}$=0.013.
The stellar mass associated to the FUV flux is then
0.25-10$\times$10$^7$\,{\msun}. 
Predicted properties based on the best FUV model for Mrk\,153
are listed in Table~\ref{predictions}.
\citet{maz93} have measured the H$\alpha$ flux in a narrow 1\arcsec\ 
slit. They obtained 9.9$\times$10$^{-14}$\,{\ergs} in total for
a group of clusters. This is larger than the value of
0.16-2.5$\times$10$^{-14}$\,{\ergs} predicted for the
large FUSE aperture which includes these clusters.


{\bf I\,Zw\,18} is a blue compact dwarf (BCD) well known for its very low 
metallicity \citep[e.g.][]{thuan83,kun95,hun95,izo97}. A review of this 
particular object was presented by \citet{kun00}. The FUSE spectrum 
of I\,Zw\,18 is shown in Figure~\ref{figlowz} with a good S/N of $\sim$18.
The stellar lines show no sign of wind profiles and they are rather faint, 
suggesting a relatively old and/or metal-poor stellar population. A 
synthetic spectrum for an instantaneous burst of 7.0\,Myr at 0.2\,{\zsun} is 
superimposed on the FUSE spectrum in Figure~\ref{figlowz}. The age of 
7\,Myr is the lower limit imposed by the lack of stars cooler than B0 in our 
spectral library at 0.2\,{\zsun} (see \S\ref{lavalsb}). With this limitation, we 
simply adopted an IMF with $\alpha$=2.2, as suggested by Aloisi, 
Tosi \& Greggio (1999). As shown in Figure~\ref{figlowz}, the model does 
not reproduce well either the {\pv} or {\ciii} line strengths. It is suspected that an 
older model of 10-20\,Myr, as found by \citet{aloi99} and \citet{ost00}, 
would not show significant change in the line strengths. However, a 
lower metallicity model is likely to produce weaker lines. This would be
consistent with the very low metallicity estimated for I\,Zw\,18 
\citep[e.g.][]{duf88,izo99,alo03}.

Adopting a model of 15\,Myr with $\alpha$=2.2 at 0.2\,\zsun, we 
can roughly estimate some physical parameters for I\,Zw\,18 from the 
measured FUV flux level. We obtain E(B$-$V)$_i$=0.0 if the Galactic 
extinction is 0.032 (NED). This low internal extinction is compatible with other 
values found in the literature \citep[e.g.][]{aloi99,can02} using optical data. 
Correcting the FUSE spectrum with this extinction, we obtain a 
stellar mass of 10$^6$-10$^7$\,{\msun} for the young burst 
while \citet{kun00} estimate it to 5$\times$10$^6$\,{\msun} from optical 
imaging with a similar aperture. 

In general, the fluxes calculated by {\tt LavalSB} are in reasonably good agreement 
with observations considering the various aperture sizes \citep{izo97,pag92}.
Nevertheless, some studies show clear indications of very young stars in a few
clusters. For example, \citet{izo97} detected a weak WR signature in an isolated 
knot (1.5\arcsec\ slit) for which they estimated 1100 O stars 
and 22 WR stars. \citet{mas99} also calculated a WR/O ratio of 0.0003. 
\citet{pag92} measured a larger EW(H$\alpha$) of 518 and 1218\AA\ for the two 
brightest knots, which is much higher that the value predicted from the FUSE data.
With the large aperture used with FUSE, it is likely that the flux from small events
is quickly diluted by the dominant 15\,Myr old stellar population for which O and
WR stars are not predicted.


{\bf NGC 1705} is an irregular BCD galaxy relatively well studied because 
of its proximity. The LWRS/FUSE aperture was used to obtain a FUV
spectrogram with S/N$\simeq$20 (see Fig.~\ref{figlowz}) for the central 
744$\times$744\,pc$^2$. From the line depths, we conclude that the metallicity is 
around 0.4\,{\zsun}. The absence of wind profile in the {\ciii} feature 
indicates a population dominated by B-type stars. Again, as it was the case 
for I\,Zw\,18, the synthesis of NGC\,1705 is limited by the lack of B-type 
stars in the spectral library at sub-solar metallicities (see \S\ref{lavalsb})
and we can only conclude that the age of this stellar population is older than
7\,Myr (see Fig.~\ref{figlowz}). This is consistent with the study of 
\citet{vaz04} where they derived an age of 12$^{+3}_{-1}$\,Myr for the 
brightest super-star cluster (NGC\,1705-1) using HST/STIS. This also 
suggests that NGC\,1705-1 may be an important contributor to the UV flux.
\citet{tosi01} resolved several individual stars using HST/WFPC2$+$NICMOS 
data and concluded that the youngest burst is 10-20\,Myr old. 
\citet{hec97b} arrived at the same result with HST/GHRS data. 
\citep{stor94} also found a low metallicity of 0.46\,{\zsun} for NGC\,1705.

Adopting a 15\,Myr instantaneous burst at 0.4\,{\zsun} with a standard IMF, 
we find E(B$-$V)$_i$=0.01$\pm$0.01 for a Galactic extinction of 0.008 (NED), 
and a stellar mass of 2.0-3.2$\times$10$^6$\,{\msun}. Along with the 
model predictions given in Table~\ref{predictions}, these are rough estimates 
as we are considering an approximate model for the lines. Nevertheless, the
predicted flux at 5500\AA\ is in good agreement with the observation 
of \citet{stor95}. However, the H$\alpha$ flux of 4.3$\times$10$^{-13}$ {\ergs} 
measured by these authors, through a 10\arcsec$\times$10\arcsec\
aperture, is a factor of 5-10 higher than the FUV prediction. \citet{anni03}
have isolated a 2-3\,Myr old population in the central 100\,pc. This
isolated population is possibly responsible for the large H$\alpha$ flux,
but its FUV flux is obviously diluted in a large aperture.

\subsection{Non-Standard IMF or Bursts Superposition}
\label{bstarimf}


{\bf NGC\,4449} is an irregular galaxy with a large number of young star 
clusters \citep[e.g. Gelatt, Hunter \& Gallagher 2001;][]{bok01}. FUSE 
obtained data for two different regions in NGC\,4449, one centered on 
the nucleus and the other on an {\hii} region located South-West from 
the nucleus. The FUV stellar lines seen (Fig.~\ref{imffig}) for the nuclear 
region are better reproduced with an instantaneous burst of 10$\pm$1\,Myr 
at {\zsun}. No continuous burst model can fit the observations. An IMF slope 
of 3.3 is required, especially for the {\ciii} structure. A 8\,Myr burst with 
$\alpha$=2.8 is slightly better for the {\pv} line profiles, but gives unsatisfying 
results for the {\ciii} feature. The {\hii} region spectrum is relatively well 
reproduced by a 5.0$\pm$0.5\,Myr burst at {\zsun}. As for the nuclear region, 
a steep IMF with $\alpha$=3.3 offers the best compromise for the {\ciii} and 
{\pv} lines.

The fact that we are clearly finding a steep IMF slope for this galaxy may be 
surprising at first since this object is known to display a standard IMF
\citep[e.g.][]{mass98,krou04}. However, the IMF slope given by the FUV line 
synthesis may be interpreted in other ways than a real steep IMF for the
massive stars. An hypothesis to explain a steep IMF and the problem related 
to {\pv} line profile fitting would be the contribution of an older stellar population
dominated by B-type stars. The FUV range (900\AA\,$<\lambda<$\,1200\AA) 
is particularly sensitive to the relative number of O and B~stars since 
they are the only stars to produce photons below 1200\AA. Considering this 
fact, if an important population dominated by B\,stars is superposed,
in the right proportion, to a younger burst, it will increase the total number of 
B~stars seen in the FUV range and then 
modify the global IMF slope. This would explain the steep IMF 
without referring to some peculiar star formation processes.

To test the hypothesis of a stellar population superposition, we 
combined synthetic spectra of two stellar 
populations with the same IMF slope of 2.35, but having different ages and 
various flux proportions. These new models have been compared to the 
single population with $\alpha$=3.3, as observed in NGC\,4449. Results 
are reported in Figure~\ref{figdilution} for young instantaneous bursts of 10 and 
5\,Myr (for the nuclear and the {\hii} regions respectively), both combined with an 
older burst of 20\,Myr. It shows that the line profiles of the young burst are, as 
expected, diluted by a B~star population of 20\,Myr. If 25 
to 50\% of the total stellar flux contributing to the light comes 
from the 20\,Myr burst, the FUV line profiles then correspond well to 
the one of a single young burst with $\alpha$=3.3. From this, we
conclude that the FUSE spectra of NGC\,4449 are produced by more than one
important stellar population, each with a standard IMF. This conclusion reinforces 
the results of \citet{gel01} who claimed that an active star formation took 
place in NGC\,4449 during the last 10\,Myr to 1\,Gyr.

We also considered the time-dependent dust obscuration effect as proposed 
by \citet{lei02b}. In this particular case, the contribution of dust being less 
pronounced, if not absent, around older stars, one can expect to observe more 
B-type than O-type stars. Following these authors, the time-dependent dust 
obscuration effect can be mixed up with a steep IMF. But for this effect to be 
seen, a continuous star formation burst must be considered, which is not 
the case for NGC\,4449.

Using Balmer line intensities, \citet{bok01} estimated that the stellar population 
in the nucleus is about 6-10\,Myr. From HST/WFPC2 and infrared spectra, 
\citet{gel01} obtained for the nucleus an age of 8-15\,Myr, and for some stellar 
clusters an age between 10\,Myr and 1\,Gyr. An age of 6.4\,Myr was estimated 
by {\citet{hill98} for the {\hii} region. All these studies are consistent with the FUV 
spectral synthesis. The metallicity of NGC\,4449 was 
estimated to be 12+log[O/H]=8.4, i.e. 1/2\,{\zsun}, according to \citet{hec98}.

Adopting a single burst model for both the nucleus and the {\hii} region, we 
find that the internal extinction is 0.01$\pm$0.01, using 
E(B$-$V)$_{Gal}$=0.019 (NED). The continuum is very blue and difficult to 
fit properly. This is consistent with the presence of an underlying older 
population contributing to the FUV spectra, in addition to a younger one. 
The corrected FUV fluxes then give stellar masses of 
1.3-2.0$\times$10$^6$\,{\msun} for each of the nucleus and the {\hii} 
region. \citet{gel01} and \citet{bok01} obtained stellar 
masses of 2 and 4$\times$10$^5$\,{\msun} for the nuclear region, which 
is a factor of two lower than the FUV result. However, \citet{bok01} used a 
2\arcsec$\times$2\arcsec\ aperture, significantly smaller 
than the LWRS/FUSE aperture, which can explain the difference.
{\citet{hill98} used a large 0.39\arcmin\ circular aperture to observe
the H$\alpha$ flux in the {\hii} region. Their result, 3.2$\times$10$^{-12}$ \,{\ergs}
is very similar to the values predicted by a single population.
Other predicted values are given in 
Table~\ref{predictions}. It is rather difficult to compare them to 
published data because of large differences between apertures.


{\bf NGC\,1140} is an irregular BCD containing numerous {\hii} regions 
\citep{hod83}. A fair quality spectrum (S/N$\simeq$8; see Fig.~\ref{imffig}) 
has been obtained with FUSE. The line depth of {\ciii} and {\pv} are 
better reproduced by a model at {\zsun}. The faint P\,Cygni 
absorption seen in the blue wing of {\ciii} is better reproduced wtih
a population of 5.0\,Myr, while the {\pv} doublet favors a population of 
6.0\,Myr. A compromise for the two indicators is a 5.0$\pm$1.0\,Myr 
instantaneous burst. Note that an IMF slope of 2.8 improves the line fitting, 
especially for the {\pv} feature. 
Continuous burst models cannot reproduce the diagnostic line
profiles. The young age found with FUV data is consistent with those given
in the literature \citep[Guseva, Izotov, \& Thuan 2000;][]{bon99}. 
As for NGC\,4449, the steep IMF slope found here is interpreted as an indication 
of multiple burst superposition, all with a standard IMF slope, where one
burst is dominated by B stars. \citet{hec98} evaluated an abundance of 
12+log[O/H]=8.0, or 0.2\,{\zsun}, while it would be of 8.46, or 
0.6\,{\zsun}, according to \citet{gus00}. The last value is in better 
agreement with the line depth seen in the FUV.

By comparing the observed FUV slope continuum to the best single burst 
model with $\alpha$=2.8, we find E(B$-$V)$_i$=0.08$\pm$0.08 using
E(B$-$V)$_{Gal}$=0.038 (NED). This extinction is in good agreement with the 
optical values of 0.10 and 0.18 obtained by \citet{stor94} and \citet{cal95}, 
respectively. Correcting for dust, we estimate a stellar mass of 
5.6$\times$10$^6$ to 1.6$\times$10$^8$\,{\msun}. H$\alpha$ fluxes of 1.4 to 
1.9$\times$10$^{-12}$ {\ergs}, measured by several authors 
\citep{maz93,cal95,stor95,ros02}, are consistent with predictions reported 
in Table~\ref{predictions}. A EW(H$\alpha$) of 138.5\AA\ \citep{maz93} and 
a continuum level of 1.4$\times$10$^{-14}$ at 5500\AA\ \citep{stor95} 
also correspond relatively well to the FUV predictions taking into account
the aperture sizes. \citet{gus00} measured a much lower value for the 
H$\alpha$ flux. Based on the H$\beta$ line and the WR bump at 4686\AA,
they also estimated the presence of 15\,547 O and 461 WNL stars, which 
are in agreement with the predictions within the given uncertainties.

\subsection{Other Starbursts:  FUSE spectra with a lower signal}
\label{othersb}


{\bf NGC\,4194} is classified as a BCD galaxy. The spectrum collected by FUSE
 for this target is rather noisy (S/N$\simeq$6; see Fig.~\ref{figsn}),
especially in the {\ciii} line section where only the LiF1B segment 
contributes. Furthermore, the {\pv~$\lambda$1118} line is contaminated by 
an airglow line. Despite these observational constraints, we obtain a reliable fit 
to the stellar lines using an instantaneous burst model of 5.5$\pm$0.5\,Myr with a 
solar metallicity. An IMF slope of 2.35 is simply adopted here because of the low
S/N. A continuous burst models cannot reproduce the data. 

Few studies have been published about the stellar populations of NGC\,4194. 
\citet{cal97b} estimates its metallicity to 1.26\,{\zsun} (i.e. 12$+$log[O/H]=8.81), 
consistent with the FUV line profiles. \citet{kin93} have detected {\siiv} 
and {\civ} lines in IUE spectra, signatures of massive stars. Their 
flat UV continuum strongly suggests the presence of A-type stars. Based on
the UV flat continuum, \citet{bon99} estimate an age of 10 to 25\,Myr for the stellar 
population. From our FUV data, we are confident that a very young population 
of 5.5$\pm$0.5\,Myr is also present in NGC\,4194. With a higher S/N, it 
would have been possible to better constrain the IMF slope and maybe 
see indications of an older population (see \S\ref{bstarimf}).


{\bf IRAS\,19245-4140 (or Tol\,1924-416)} is a BCD galaxy. The observed FUV 
spectrum of IRAS\,19245-4140 is presented in Figure~\ref{figsn}, showing a S/N 
of 6. The best fit for the relatively shallow stellar line profiles is obtained for an 
instantaneous models at 0.4\,{\zsun}. An instantaneous burst with an age of 3.5 
to 6.0\,Myr is ideal to reproduce the observed {\ciii} line while an age of 4.5 to 
6.0\,Myr gives better results for the {\pv} doublet. We therefore adopt an age of 
5.0$\pm$1.0\,Myr. A standard IMF was adopted because of the low S/N. A 
continuous burst scenario is never satisfying because the observed {\ciii} line 
does not display any P\,Cygni profile and the observed {\pv} lines are too 
narrow. An age of 5\,Myr is in agreement with other works 
\citep{ber85,kin93,rai00,bon00}. Based on the flux measurements of nebular lines 
from \citet{ber85}, we calculate an oxygen abundance of 12$+$log[O/H]=8.1 
(0.25\,\zsun) for IRAS\,19245$+$4140, consistent with FUV synthesis.

From the FUV continuum slope, we find an internal extinction of E(B$-$V)=0.1$\pm$0.1, 
with a better fit if the extinction is closer to zero. This result agrees with the low reddening 
found by \citet{stor94} and \citet{cal95}. The observed FUV flux level 
leads to a stellar mass of 10$^6$-10$^8$\,{\msun} for the young 
burst. The H$\alpha$ fluxes measured by \citet{stor95} and \citet{cal95} 
are consistent with {\tt LavalSB} calculations 
(Tab.~\ref{predictions}). The F(5500) value obtained by \citet{stor95} is 
also in agreement with the predictions. \citet{rai00} also see
an old population of $\sim$100\,Myr in addition to the young one.
However the presence of the second component is impossible to establish 
from FUV data.


{\bf NGC\,3504} is a spiral galaxy composed of a young nuclear starburst and a 
non-thermal source \citep{keel84}. Figure~\ref{figsn} shows the FUSE 
spectrum of NGC\,3504 with S/N$\simeq$8. From the presence of stellar wind
features, it is clear that the nuclear burst dominates the non-thermal source and 
can easily be studied at short wavelengths. Note the presence of an airglow 
line near {\pv~$\lambda$1128}. The wind signatures are so pronounced that a model
at {2\,\zsun} and a very young age of 2.5-3.5\,Myr is required, with a 
better fit at 2.5\,Myr. As shown in Figure~\ref{figsn}, the observed stellar lines 
are not perfectly fitted by the best model due to the lack of an appropriate
spectral library at high metallicity (see~\S\ref{lavalsb}). Because of the limitation
imposed by the spectral library, we simply adopted a standard IMF slope.
Based on the line behavior of models at {\zsun}, continuous burst scenarios have 
been rejected because they cannot reproduce the strong P\,Cygni for the {\ciii} line.
Other studies on NGC\,3504 mostly focus on the non-thermal source, but the few
ones treating the galaxy stellar populations seems to be 
in good agreement with FUV synthesis \citep{sek87,ter90,kin93,elm97}. The 
galaxy is metal rich, with 2.5\,{\zsun} according to \citet{hec98}, which is consistent 
with the FUV synthesis.

Using a Galactic extinction of 0.027 (NED), we calculate that 
E(B$-$V)$_i$=0.30$\pm$0.05, leading to 
Log(M$_{\star}$)=8.8$\pm$0.5\,{\msun}. This mass is an upper limit 
since we did not consider
a contribution from the non-thermal source. Other calculated parameters are
given in Table~\ref{predictions}. However, the lack of studies on the stellar 
content of NGC\,3504 allows no more comparison.


{\bf NGC\,5194} (M\,51) is part of an interacting system with NGC\,5195, and is known 
as the Whirlpool nebula. A faint non-thermal source has also been observed in the
nucleus \citep{hec80}. Five data sets centered on various regions of 
NGC\,5194 have been obtained with FUSE. Three of these sets have a poor quality 
while the two other sets 
have S/N of $\sim$3 (B10601 and B10604), which are also individually too 
low for synthesis. The B10601 set is centered on the nucleus and the 
B10604 targets a bright {\hii} region located 2\arcmin\ North of the nucleus. 
Amazingly, although they include very different physical regions, these two 
spectra look similar with broad P\,Cygni profiles lost into the noise. Therefore,
in order to increase the S/N  and facilitate the synthesis, we decided to combine
the two data sets. A careful interpretation is then recommended as if the final 
spectrum was obtained, roughly, through a larger aperture.

The combined FUV spectrum is shown in Figure~\ref{figsn}. The resulting S/N is of 
$\sim$4, barely enough for the spectral synthesis. As for M\,83 and NGC\,3504 
(Fig.~\ref{figproto} and \ref{figsn}, respectively), the broad wind profile in {\ciii} suggests 
a young and metal rich stellar population. Considering models at {\zsun} and 2\,{\zsun}, 
and adopting a standard IMF, we can reproduce the diagnostic lines with an 
instantaneous burst of 3.0-3.5\,Myr. If we try to perform the synthesis on the individual 
sets of observation, we roughly obtained an age of about 2 and 4\,Myr for the {\hii} region 
and the nucleus, respectively. Therefore, the age obtained for the average spectrum 
seems meaningful taking into account the differences between the two regions, the low 
S/N, and the lack of a metal rich spectral library (see \S\ref{lavalsb}). The HUT spectrum 
of NGC\,5194 studied by \citet{lei02} revealed an age of $\sim$5\,Myr, in good agreement 
with the FUV synthesis. The IUE spectrum from \citep{kin93} suggests that the dominant 
stars in NGC\,5194 are A- and G-type stars. In any case, such stars are too cold to be 
detected below 1200\AA. \citet{zar94} measured an oxygen abundance of 9.2, 
corresponding to 3.1\,{\zsun}, in the central region, consistent with the 
strong FUV wind profile of {\ciii}.

Using a model of 3.5\,Myr at 2\,{\zsun} and a Galactic extinction of 0.035 (NED), 
we measure an internal extinction of 0.25$\pm$0.15, which leads to a global 
stellar mass of the order of 10$^7$\,{\msun}. Predicted values obtained 
from the combined FUSE data sets are reported in Table~\ref{predictions}. 
If these values are to be compared with observations, one must remember to
consider the global flux from the nucleus and the {\hii} region and also to take
into account the non-thermal contribution.


{\bf NGC\,3991} is a Magellanic type galaxy, member of the interacting system 
which includes NGC\,3994 and NGC\,3995 \citep{keel85}. It contains about 25
stellar clusters \citep{meu95}, probably responsible for most of the FUV flux.
Figure~\ref{figsn} shows the observed spectra obtained for NGC\,3991 
(S/N$\simeq$6). Because of the redshift, the {\ciii} structure is incomplete. 
However, the presence of an extended absorption in the blue wing of {\ciii}, 
due to evolved O~stars, is evident. Models at {\zsun} correspond well to the 
observed line strengths of {\pv}. The best-fitting for the {\pv} doublet is 
obtained at 4.0-4.5\,Myr with $\alpha$=2.8, especially to fit the  
{\pv}~$\lambda$1128 P\,Cygni profile. An instantaneous burst of 
3.5-4.0\,Myr can reproduce the {\ciii} blue wing for both IMF slopes of 2.35 
and 2.8. We then conclude that the young stellar
population of NGC\,3991 is dominated by massive stars of 4.0$\pm$0.5\,Myr 
at $\sim${\zsun} with an IMF slope of $\sim$2.35-2.8. Although the noise level 
in the FUV spectrum of NGC\,3991 is relatively high, the steep IMF slope
suggests that one or several older bursts, dominated by B stars, may contribute
to the FUV line profiles. As for NGC\,7714, it is possible to reproduce the observed 
line profiles with a continuous burst of star formation, but we suspect that this is not 
real as a result of a degeneracy effect between an instantaneous model at 4.5\,Myr 
and a continuous burst of 10-20\,Myr. According to the work of \citet{bon99}, about 
60\% of the flux at 2646\AA\ is produced by O-type stars of less than 5\,Myr, 
consistent with the FUV synthesis for an instantaneous burst. \citet{hec98} 
estimated the metallicity of NGC\,3991 to 0.6\,\zsun, indicating that the FUV models 
at {\zsun} can fit a wide range toward lower metallicities.

Adopting the instantaneous burst scenario and a E(B$-$V)$_{Gal}$=0.022 (NED), 
we estimate the internal extinction value to 0.3$\pm$0.1. This value is 
higher than the 0.07 value obtained by \citet{meu95}. The E(B$-$V)$_i$ 
implies a stellar mass of 10$^9$ to 10$^{11}$\,{\msun}. Unfortunately, 
the FUV synthetic results cannot be fully tested since very few works 
are related to the stellar populations of NGC\,3991. The observed H$\alpha$ flux 
values, uncorrected for extinction, obtained by \citet{dah85}, \citet{keel85}, and 
\citet{meu95} with various apertures are between 2 and 3$\times$10$^{-13}$\,{\ergs}, 
a factor of 10$^3$ lower than the {\tt LavalSB} calculation.


{\bf NGC\,7673 (Mrk\,325)} displays numerous aggregates \citep{ben82}.
Figure~\ref{figsn} shows the observed FUV spectrum with S/N=9, which
includes most of the known star clusters. The {\ciii} line is incomplete because 
of the galaxy redshift, but the {\ciii} blue wing do not show obvious absorption 
in comparison to NGC\,3991 (having a similar redshift). This suggests an 
older burst and/or a lower metallicity. The {\ciii} and {\pv} lines are indeed 
well reproduced by an instantaneous burst of 
6.5$\pm$0.5\,Myr at {\zsun} with $\alpha$(IMF)=2.35-2.8. Continuous models 
do not reproduce the observed {\pv} line profiles. A steep IMF slope suggests the 
superposition of an older burst dominated by B stars (\S\ref{bstarimf}),
which seems to be consistent with the IUE spectrum of \citet{kin93} where an 
important contribution from B-type stars is observed. \citet{bon99} also concluded 
that 50\% of the UV flux comes from OB stars. \citep{hec98} give a metallicity of 
0.6\,{\zsun} for NGC\,7673, which is consistent with the range of metallicity 
covered by the {\zsun} model (see also NGC\,1140 in \S\ref{bstarimf}).

Adopting a burst of 6.5\,Myr at {\zsun} and a standard IMF, we estimate an 
internal extinction of 0.07$\pm$0.07, with a better fit closer to zero, using E(B$-$V)$_{Gal}$=0.043 
(NED). This value is much below the E(B$-$V)$_i$ of 0.42 
obtained by \citet{stor94} from the Balmer decrement through a 
10\arcsec$\times$20\arcsec. Although the Balmer decrement usually gives 
twice the extinction values based on stellar indicators, this is still high 
compared to the FUV result. However the low extinction gives a 
stellar mass of $\sim$10$^7$\,{\msun} which is 
consistent to the 9$\times$10$^7$\,{\msun} estimated by \citet{duf82} 
through a 40\arcsec$\times$40\arcsec\ aperture. The predicted 
H$\alpha$ and F(5500) fluxes (see Tab.~\ref{predictions}) are in 
agreement with the observations of McQuade, Calzetti \& Kinney 
(1995; circular aperture of 13.5\arcsec).  An EW(H$\alpha$) of 106\AA\ 
measured by \citet{keel92} is in good agreement with the FUV predictions.


{\bf Mrk\,054} is a very blue starburst galaxy \citep{gon84,kin93} and one of the most 
distant object of the sample ($z$=0.045). Because of the large redshift, the {\ciii} 
line is outside of the FUSE wavelength range (see Fig.~\ref{figsn}). Furthermore,
the {\pv~$\lambda$1118} line is contaminated by an airglow emission. Fortunately, 
the S/N ratio of the FUSE spectrum is good enough and the {\pv} lines can still be used
for the synthesis. The {\pv} line profiles suggest a metallicity around solar and 
an age of (10$\pm$1)\,Myr. A standard IMF has been assumed. There is no work 
in the literature that can help us confirm the metallicity of Mrk\,054. \citet{bon00} 
estimate that 54\% of the flux at 2646\AA\ from IUE data comes from a young 
stellar population of $\sim$2.5\,Myr. 

With a 10\,Myr burst at {\zsun} and a Galactic extinction of 0.015 (NED),
we find an internal extinction of 0.1$\pm$0.1, with a better fit closer to zero. 
It corresponds to the estimation of \citet{bon00}. This value leads to a stellar 
mass of $\sim$10$^8$\,{\msun}. Model predictions are listed
in Table~\ref{predictions}. \citet{mae87} obtained EW(H$\alpha$)=40\AA, 
which is quite high compared to the {\tt LavalSB} calculation of 11$\pm$6\AA, 
but the uncertainty on the result of these authors may be large as they used 
photographic plates.


{\bf Tol\,0440$-$381} is the second most distant object of our sample 
($z$=0.041). Because of the redshift, the {\ciii} line is not observed 
(see Fig.~\ref{figsn}). The {\pv~$\lambda$1118} line is dominated by noise. 
Synthesis is then restricted to the {\pv~$\lambda$1128} line. Models at 
{\zsun} are good to reproduce the line depth, but the absence of the {\ciii} 
feature combined with the noise does not allow to really distinguish between 
the solar and sub-solar metallicities. The best fit is obtained 
for an instantaneous burst of 5.0\,Myr, while ages between 4.5 to 6.0\,Myr 
are suitable. A standard IMF has been assumed. A few studies on 
Tol\,0440$-$381 focus on stellar populations \citep{maseg91,rai00}. They
are usually very consisten with the FUV synthesis since they report the presence of WR
signatures. From nebular line intensities obtained by \citet{ter91}, we calculated 
an oxygen abundance of 8.2 (0.3\,{\zsun}), in agreement with \citet{den02}.
This favors sub-solar metallicity models for the FUV synthesis. 

Using E(B$-$V)$_{Gal}$=0.015 (NED), we obtain that E(B$-$V)$_i$=0.20$\pm$0.15, 
similar to the nebular value of 0.24 obtained by \citet{cer94}. This extinction gives a 
stellar mass around 10$^{9}$\,{\msun} for the burst. {\bf As for IRAS\,08339$+$6517, 
the stellar mass value of Tol\,0440$-$381 shows large uncertainties due to a shorter 
wavelength coverage and large uncertainties on the extinction value. This mass value 
must be interpreted with care.} The optical values predicted from the FUV information 
are in relative good agreement with the observations of \citet{cam86} and \citet{ter91} 
considering the differences in aperture size and uncertainties.

\subsection{Young Stellar Populations in Seyfert Galaxies}
\label{synsyft}


{\bf NGC\,1672} shows a Seyfert\,2 type nuclear activity \citep{mou89} 
as well as a nuclear ring of intense star formation (e.g. Storchi-Bergmann, 
Wilson \& Baldwin 1996). The FUSE spectrum of NGC\,1672, centered on 
the nucleus (Fig.~\ref{figagn}), displays a broad and deep {\ciii} line profile, 
suggesting the presence of evolved metal rich O-type stars. The 
{\pv~$\lambda$1118} line is contaminated by an important airglow emission. 
Models at 2\,{\zsun} are required to explain the line width and depth of {\ciii}, 
and a better fit is obtained for a 3.5$\pm$1.0\,Myr population. {\pv} lines 
can be reproduced by a model at 3.5-4.0\,Myr. Overall, the best fit is obtained for a 
3.5$\pm$1.0\,Myr instantaneous burst. A standard IMF slope has been assumed
because of the spectral library limitation at this metallicity. \citet{bon98} calculated 
that the dominating stellar population at 2646\AA\ is 10$^8$\,yr (but these older 
stars do not emit below 1200\AA) and that stars of less than 20\,Myr contribute to 
less than 20\% to the flux. \citet{gar90} synthesized the IUE spectrum of NGC\,1672 and 
obtained an age of about 8\,Myr for the burst, still older than what is suggested 
by the strong P\,Cygni profile in {\ciii}. In agreement with the FUV synthesis, 
\citet{zar94} measured an abundance 12+log[O/H]=9.1 ($\sim$2.5\,{\zsun}) and
supported by \citet{stor96,stor98} who found a similar value of 9.2.

Considering a Galactic extinction of 0.023 (NED), we find an internal 
extinction of 0.15$\pm$0.10. This is relatively consistent with the extinction 
of 0.6 obtained by \citet{stor94} with the Balmer decrement, considering the 
uncertainties. The mass of the young stellar population estimated from the 
FUV flux is about 10$^7$\,{\msun}, which is an upper limit considering 
the non-thermal source. However, according to \citet{hec95}, a maximum of 
20\% of the flux at 1500\AA\ in Seyfert\,2 nuclei comes from the non-thermal 
source. This contribution seems to be generally lower than 10\% 
\citep{hec97a,gonz98b}. The optical flux values reported in 
Table~\ref{predictions} are higher than the observations of \citet{stor95}. 
This can be explained by the presence of the non-stellar activity. Underlying 
stellar populations could also contribute to the optical flux.


{\bf NGC\,7496} is a barred spiral galaxy with a Seyfert\,2 nucleus. The 
nuclear region also displays many {\hii} regions \citep{veron81}, which are 
responsible for an intense far-infrared luminosity. The FUSE spectrum 
shows strong evidence for such OB~stars (Fig.~\ref{figagn}) with a well-developed 
wind profile in {\ciii}. An airglow line is superposed to the {\pv~$\lambda$1128} 
line. The comparison with FUV synthetic spectra suggests a metallicity higher 
than solar, and an age of 2.5$\pm$0.5\,Myr, assuming a standard IMF. This age 
is consistent with the result of \citet{bon98}. The metallicity of the nucleus is known 
to be around 2\,{\zsun} \citep[12+log{[O/H]}=9.0;][]{hec98}, also in agreement with 
the FUV synthesis.

Comparing the observed and synthetic FUV continuum slopes, we obtain 
that E(B$-$V)$_i$=0.03$\pm$0.02, with a better fit closer to zero, and using a Galactic extinction of 0.01 (NED). This 
extinction is much lower than the 0.6 value obtained by \citet{stor94}. The FUV 
flux implies a stellar mass of 3-10$\times$10$^5$\,{\msun}, which is an upper limit
considering that the non-thermal source is also contributing to the FUV flux. 
\citet{stor95} measured F(H$\alpha$)=5.8$\times$10$^{-13}$ {\ergs} consistent with 
the prediction, but they also obtained 
F(5500)=9.9$\times$10$^{-15}$ {\ergs}, which is slightly higher than the 
predictions reported in Table~\ref{predictions}. The difference can be explained by the
non-thermal activity and/or underlying older stellar populations.


{\bf NGC\,1667} is a Seyfert\,2 galaxy where several {\hii} regions are observed 
along the spiral arms and in a nuclear ring \citep{gonz97b}. \citet{thuan84} studied 
the IUE spectrum of NGC\,1667 and concluded that the active nucleus is lying in an 
important starburst. As shown in Figure~\ref{figagn}, although the {\ciii} line is
missing because of the redshift, stellar lines of {\pv} are clearly observed.
The {\ovi} doublet (not shown) also displays a P\,Cygni profile, which confirms
the young stellar population contribution in the FUV range. The {\pv} line 
depth suggests a metal rich population at {\zsun} or 2\,{\zsun}, but it is 
impossible to distinguish between the two metallicities without the {\ciii} line.
Adopting a standard IMF, an age of 5.0$\pm$0.5,Myr 
is obtained if the metallicity is {\zsun}, or 4.0$\pm$1.0\,Myr if we 
use a 2\,{\zsun} model. These ages are consistent with previous works
\citep{thuan84,bon98}. \citet{stor90} measured the nitrogen and sulfur 
abundances of NGC\,1667 and, taking into account the contamination by the 
active nucleus, they obtained a high metallicity around 3\,{\zsun}.

Considering the observed metallicity, we adopt the model at 4.0\,Myr and
2\,{\zsun} for NGC\,1667, and we find an internal extinction of 
0.08$\pm$0.08 using a E(B$-$V)$_{Gal}$=0.0. We calculated that
Log(M$_{\star}$)=7.5$^{+0.5}_{-1}$, which is only an upper limit since the
AGN also contribute to the flux measured by FUSE. Predicted fluxes in the
visible range (Tab.~\ref{predictions}) are also higher limits. The predicted values 
are in agreement with the observations \citep{mcq95,stor95,ho97,thuan84}, 
considering the uncertainties.


{\bf NGC\,1068 (M\,77)} is a nearby spiral galaxy ($z$=0.0038) well-studied
for its Seyfert\,2 nucleus and its nuclear ring of star formation \citep[e.g.][]{ant85}. 
The FUSE spectrum of the central 30\arcsec$\times$30\arcsec\ is presented in 
Figure~\ref{figagn}. It displays broad and extremely shallow absorption features 
at wavelengths corresponding to the stellar lines. 
However, no model for a young stellar population can reproduce these unusual 
profiles. In Figure~\ref{figagn}, the synthetic spectrum for an instantaneous burst 
of 5\,Myr at {\zsun} with a standard IMF slope is shown superposed to the 
observation for a simple comparison. It seems here that very massive stars may 
contribute to the spectrum, but they are strongly diluted by an additional continuum, 
likely from the active nucleus. However, our attempt to include a featureless 
power-law continuum to the synthetic spectrum of a stellar population was not 
successful.

\subsection{Seyfert Galaxies Dominated by Non-Stellar Activity}
\label{syft1}


{\bf NGC\,4151} is classified as a Seyfert\,1.5 galaxy because of its remarkable 
flux variabilities between the Seyfert\,1 and 2 types. During its low activity phase, 
absorption features are observed in the FUV (Fig.~\ref{figagn}). A closer look at 
these lines reveals that they are blueshifted by 2-3\AA\ compared to the synthetic stellar 
features of a 10\,Myr burst at {\zsun}. The FUSE spectrum was
corrected for the radial velocity of 995 km\,s$^{-1}$ as given by NED. 
We also noted that the {\ovi} feature
do show non-stellar features in emission. Despite the fact that the observed line
profiles in NGC\,4151 may mimic the usual diagnostic lines of young starbursts,
we are actually looking at signatures from hot gas located very near 
the central engine, ionized, and pushed away from it as shown by e.g.
\citet{kriss92}, \citet{espey98}, and \citet{cren99}. This explain the shifted
features and consequently no signature 
of massive stars are detected within the FUSE aperture.


{\bf NGC\,3783, NGC\,7469, and NGC\,5548} are Seyfert\,1 galaxies for which 
FUSE obtained a spectrum with a good S/N (see the example of NGC\,3783 in 
Fig.~\ref{figagn}}).  As expected, the FUV spectrum of these three Seyfert\,1 nuclei 
are entirely dominated by the non-thermal activity and no obvious absorption 
features, other than interstellar lines or detector defects, was observed. At shorter 
wavelengths,the FUSE spectra of these objects \citep[see e.g.][]{kri03} display a 
flat continuum with broad emission lines for e.g. {\ciii~$\lambda$977}, 
{\niii~$\lambda$991}, the {\ovi~$\lambda\lambda$1032, 1038} doublet, and the 
{\heii~$\lambda$1085} line. These objects are dominated by the non-thermal
source and, if present, signature of massive stars are completely diluted.

\section{Conclusion}
\label{conclusion}

In this work, we characterized the physical properties of very young  stellar
populations contained in 24 local starbursts. FUSE data and  evolutionary
spectral synthesis with {\tt LavalSB} allowed us to obtain accurate  ages,
estimates of stellar masses and internal extinctions.
We also calculated and compared values for several parameters in the  visible range
produced by the young stellar populations observed in the FUV.

While the metallicity cannot be evaluated with a great precision, we  were able to
distinguish between 4 metallicity ranges corresponding to the 4  evolutionary track
metallicities used by {\tt LavalSB}. Unlike the UV wavelength range  above 1200\AA, 
we were able to distinguish between solar metallicity and richer populations  due to
unsaturated {\ciii} line profile. In a few cases where the FUV synthesis 
indicated a solar metallicity  (NGC\,1140, NGC\,4449, NGC\,7673), we found 
that the literature was sometimes proposing a lower value, but still higher than the
LMC metallicity.  We therefore conclude that the FUV line profiles may be less
sensitive to the metallicity between values of $\sim${\zsun} and 0.4\,{\zsun}.  Although our
low metallicity models are limited by the lack of B-type stars, we were able to
clearly identify low metallicity stellar populations and to estimate a lower age limit 
(i.e. NGC\,1705, I\,Zw\,18).

From the FUV point of view, star formation in starbursts seems to occur on a very
short period of time, favoring the instantaneous burst scenario.
In most cases, continuous star formation models obviously did not represent
the observed situations. The predominance of instantaneous bursts in our sample of
starburst galaxies is likely the result of an observational bias. The bias is due to
the continuum flux level of individual O and B stars below 1200A, which varies
significantly from one spectral type to another. The effect is present above 1200A,
but is more dramatic below 1200A because the SED of OB stars in the FUV is coincident
with the peak of their black body radiation spectrum. A small variation of temperature will
affect significantly more the bluer side of the black body radiation spectrum than
the redder side. Consequently, any property derived from stellar line profiles in the
FUV range is biased toward the young populations. This idea is supported by the work
of \citet{bruw03} where they compared the UV line profiles
of individual massive stars observed in NGC604 with those of the integrated spectrum.
They found that a very small number (10) of very bright stars dominate the line profiles
of the whole stellar cluster. When looking at starburst galaxies in the FUV with a large
aperture, it should not be surprising to see that young populations of a few Myr
preferably show up and that the stellar line profiles are in better agreement with
instantaneous burst models.

For spectra having a sufficient S/N, we constrained the IMF slope using the
sensitivity of line profiles. In general, the IMF slope $\alpha $ is near 2.35.
We used this property to detect underlying stellar populations dominated
by B-type stars in NGC\,4449, NGC\,3991, and NGC\,1140, probably  resulting from
one or several previous burst event a few 10$^7$ to 10$^8$ years ago.

We also detected very young stellar populations of less than 5\,Myr in
Seyfert\,2 nuclei. However, no sign of such population has been detected
in Seyfert\,1 nuclei, the central engine being dominant in flux.

\section*{Acknowledgments}

The authors thanks C. Leitherer for comments on the manuscript. 
This work was supported by the Natural Sciences and Engineering Research Council
of Canada and by the Fonds FCAR of the Government of Qu\'ebec. 
This research has made use of the NASA/IPAC Extragalactic Database (NED) 
which is operated by the Jet Propulsion Laboratory, 
California Institute of Technology, under contract with the NASA.

\begin{table*}
\centering
\begin{minipage}{140mm}
\caption{Sample of Galaxies Observed with FUSE}
\label{echantillon}
\begin{tabular}{@{}llcclrllr@{}}
\hline
Name         &Morph/Activity      &RA(J2000)   &DEC(J2000)    
&E(B$-$V)$_{Gal}$\footnote{From NED if nothing else is specified.} &D\footnote{Distances from \citet{hec98} 
and other authors cited in the text.} 
& FUSE ID &Exp. \\
             &                    &($^h$\,$^m$\,$^s$)     &($^{\circ}$ \arcmin\  \arcsec) 
& &[Mpc] &  &[sec] \\
\hline
IRAS\,08339$+$6517 &SB/starburst  &08:38:23.20 &$+$65:07:16.0 &0.092 &78.0  &B00401 &35162 \\
IRAS\,19245-4140  &pec/Starburst  &19:27:58.02 &$-$41:34:27.7 &0.00\footnote{See text.} &36.5  &A02306 &4767  \\
IZw\,18      &BCD                 &09:34:02.30 &$+$55:14:25.0 &0.00$^c$ &14.3  &P19801 &31647  \\
M\,83        &SAB(s)c/Starburst   &13:37:00.51 &$-$29:52:00.5 &0.066 &3.8   &A04605 &26528  \\
MRK\,054     &Sc/Starburst        &12:56:55.90 &$+$32:26:52.0 &0.015 &180.0 &A05201 &25177  \\
MRK\,153     &Sc/Starburst        &10:49:05.15 &$+$52:20:05.1 &0.00$^c$ &37.1  &A09401 &103724 \\
NGC\,1068    &SA(rs)b/Syft\,1-2   &02:42:40.50 &$-$00:00:51.7 &0.034 &      &A13902 &75987  \\
	     &                    &02:42:40.72 &$-$00:00:48.5 &      &      &P11102 &22579  \\
NGC\,1140    &IBm\,pec/Syft\,2    &02:54:33.58 &$-$10:01:39.9 &0.038 &19.6  &A02305 &4017   \\
NGC\,1667    &SAB(r)c/Syft\,2     &04:48:37.14 &$-$06:19:11.9 &0.00$^c$ &60.0  &C04901 &27651  \\
NGC\,1672    &SB(r)bc/Syft\,2     &04:45:42.68 &$-$59:14:49.7 &0.023 &15.5  &C10901 &11883  \\
NGC\,1705    &SA0\,pec/Starburst  &04:54:13.48 &$-$53:21:39.4 &0.008 &5.7   &A04601 &24026  \\
NGC\,3310    &SAB(r)bc/Starburst  &10:38:45.69 &$+$53:30:06.0 &0.022 &17.9  &A04602 &26952  \\
NGC\,3504    &SAB(s)ab/Starburst  &11:03:11.21 &$+$27:58:21.0 &0.027 &25.4  &A02301 &13455  \\
NGC\,3690    &SB\,pec/{\hii}    &11:28:31.00 &$+$58:33:41.0 &0.017 &45.0  &B00402 &59030  \\
NGC\,3991    &Sm/Starburst        &11:57:30.80 &$+$32:20:12.1 &0.022 &46.3  &A02302 &6842   \\
NGC\,4151    &SAB(rs)ab/Syft\,1.5 &12:10:32.58 &$+$39:24:20.6 &0.028 &      &C09201 &12916  \\
	     &                    &12:10:32.51 &$+$39:24:20.8 &      &      &P11105 &20638  \\
	     &                    &12:10:32.60 &$+$39:24:21.0 &      &      &P2110201 &14525 \\
             &                    &            &              &      &      &P2110202 &6601 \\
NGC\,4194    &IBm\,pec/{\hii}     &12:14:09.70 &$+$54:31:38.0 &0.015 &38.2  &C04803 &87236  \\
NGC\,4214    &IAB(s)m/Starburst   &12:15:39.41 &$+$36:19:35.1 &0.022 &3.4   &A04603 &17981  \\
NGC\,4449-Nuc &IBm/Starburst      &12:28:10.81 &$+$44:05:42.9 &0.00$^c$ &3.1   &A04606 &8366   \\
NGC\,4449-HII &IBm/{\hii}         &12:28:09.37 &$+$44:05:15.8 &      &3.1   &A04607 &14042  \\
NGC\,5194    &SA(s)bc\,pec/{\hii} &13:29:52.35 &$+$47:11:43.8 &0.035 &7.7   &B10601 &4473   \\
	     &                    &13:29:59.17 &$+$47:13:53.9 &      &7.7   &B10604 &3600   \\
NGC\,5253    &Im\,pec/Starburst   &13:39:55.97 &$-$31:38:27.0 &0.017$^c$ &4.1   &A04604 &27344  \\
NGC\,7496    &SB(rs)bc/Syft\,2    &23:09:47.27 &$-$43:25:40.0 &0.010 &20.5  &P10741 &13305  \\
NGC\,7673    &SAc\,pec/Starburst  &23:27:41.59 &$+$23:35:30.7 &0.00$^c$ &47.3  &A02303 &9930   \\
NGC\,7714 &SB(s)b\,pec/Starburst  &23:36:14.10 &$+$02:09:18.6 &0.08\footnote{\citet{gonz99}} &37.5  &A02304 &8678   \\
	  &                       &23:36:14.00 &$+$02:09:19.0 &      &37.5  &A08606 &5970   \\
          &                       &23:36:14.10 &$+$02:09:19.0 &      &37.5  &B00403 &16130  \\
TOL\,0440-381 &BCD/WR+{\hii}      &04:42:08.03 &$-$38:01:10.8 &0.015 &164.0 &A05202 &34411  \\
\hline
\end{tabular}
\end{minipage}
\end{table*}

\begin{table*}
 \centering
 \begin{minipage}{75mm}
  \caption{Spectral Bands for Continuum Slope $\beta$}
  \label{slopebands}
  \begin{tabular}{@{}ccc@{}}
  \hline
Band & Log $\lambda$ & $\lambda$ [\AA] \\
  \hline
  1 & 3.024 - 3.025 & 1056.818 - 1059.254 \\
  2 & 3.033 - 3.038 & 1078.947 - 1091.440 \\
  3 & 3.042 - 3.044 & 1102.539 - 1106.624 \\
  4 & 3.047 - 3.048 & 1114.295 - 1116.863 \\
  5 & 3.059 - 3.066 & 1145.513 - 1164.126 \\
\hline
\end{tabular}
\end{minipage}
\end{table*}

\begin{table*}
 \centering
 \begin{minipage}{140mm}
 \caption{Physical Parameters of Young Stellar Populations obtained from FUV Spectral Synthesis}
 \label{allsyn} 
 \begin{tabular}{@{}lclrrcrc@{}}
 \hline
Galaxy & Age  & $\alpha$ & Z$_{code}$ & $\beta_{obs}$\footnote{FUV Continuum slope measured after correction for Galactic extinction.} & E(B$-$V)$_i$ & Log(M$_{\star}$)\footnote{{\bf Stellar mass values derived from FUV spectra are strongly dependent on data quality and extinction. See text for details.} }  & F(1150)\footnote{Corrected for Galactic extinction only. Flux in {\ergs}.} \\
       & [Myr] &         &       &     &  & [\msun] &  [$\times$10$^{-14}$]\\
\hline
IRAS\,08339$+$6517 & 7.0$\pm$0.3 & 2.35 & {\zsun} & 3$\pm$2 & 0.30$\pm$0.15 & 10$\pm$2 & 7.9$\pm$0.8 \\
IRAS\,19245$+$4140 & 5.0$\pm$1.0 & 2.35\footnote{Standard IMF slope ($\alpha$=2.35) adopted due to S/N or other. See text for details. } & 0.4{\zsun} & $-$4$\pm$1 & 0.1$\pm$0.1 & 6$^{+2}_{-0}$ & 6$\pm$2 \\
IZw\,18       &$>$7 & 2.20\footnote{IMF slope adopted from \citet{aloi99}.} & $<$0.2{\zsun} & $-$2.1$\pm$0.9 & 0.0$\pm$0.1 & 6$^{+1}_{-0}$ & 2$\pm$1 \\
M\,83         &  3.5$\pm$0.5 & 2.35$^c$ & 2{\zsun} & $-$1.0$\pm$0.7 & 0.08$\pm$0.08 & 6.2$\pm$0.8 & 60$\pm$6 \\
Mrk\,054      & 10$\pm$1 & 2.35$^c$ & {\zsun} & $-$1$\pm$2 & 0.1$\pm$0.1 & 8$^{+2}_{-0}$ & 0.5$\pm$0.3\footnote{Extrapolated value because of the redshift.} \\
Mrk\,153      &  6.5$\pm$1.0 & 2.35$^c$ & 0.2{\zsun} & $-$2.8$\pm$0.5 & 0.07$\pm$0.05 & 7.2$\pm$0.6 & 4.2$\pm$0.4 \\
NGC\,1140     &  5.0$\pm$1.0 & 2.80 & {\zsun} & $-$1.1$\pm$0.5 & 0.08$\pm$0.08 & 7.5$\pm$0.8 & 14$\pm$1 \\
NGC\,1667     &  4.0$\pm$1.0 & 2.35$^c$ & 2{\zsun} & $-$0.9$\pm$0.9 & 0.08$\pm$0.08 & $<$7$\pm$1 & 1.0$\pm$0.5 \\
NGC\,1672     &  3.5$\pm$1.0 & 2.35$^c$ & 2{\zsun} & 0$\pm$1 & 0.15$\pm$0.10 & $<$7$\pm$1 & 4.3$\pm$0.4 \\
NGC\,1705     &$>$7.5& 2.35$^c$ & 0.4{\zsun} & $-$3.2$\pm$0.2 & 0.01$\pm$0.01 & 6.3$^{+0.2}_{-0.0}$ & 40$\pm$4\\
NGC\,3310     & 18$\pm$2 & 2.35 & 2{\zsun} &  0.3$\pm$0.3 & 0.05$\pm$0.05 & 8.1$\pm$0.5 & 40$\pm$6 \\
NGC\,3504     &  2.5$^{+1}_{-0}$ & 2.35$^c$ & 2{\zsun} & 1.1$\pm$0.4  & 0.30$\pm$0.05 & 8.8$\pm$0.5 & 4.9$\pm$0.5\\
NGC\,3690     &  6.5$\pm$0.5 & 2.35 & {\zsun} & $-$0.3$\pm$0.4 & 0.15$\pm$0.05 & 8.5$\pm$0.5 & 7.0$\pm$0.7 \\
NGC\,3991     &  4.0$\pm$0.5 & 2.80 & {\zsun} & 1.1$\pm$0.9  & 0.3$\pm$0.1 & 10$\pm$1 & 14$\pm$1 \\
NGC\,4194     &  5.5$\pm$0.5 & 2.35$^c$ & {\zsun} & 3$\pm$1 & 0.4$\pm$0.1 & 10$\pm$1 & 1.6$\pm$0.2 \\
NGC\,4214     &  5.5$\pm$1.0 & 2.35$^c$ & 0.4{\zsun} & $-$0.4$\pm$0.3 & 0.13$\pm$0.03 & 6.3$\pm$0.3 & 35$\pm$9 \\
NGC\,4449-Nuc & 10$\pm$1 & 3.30 & {\zsun} & $-$2.7$\pm$0.2 & 0.01$\pm$0.01 & 6.2$\pm$0.1 & 22$\pm$2 \\
NGC\,4449-\hii\ &  5.0$\pm$0.5 & 3.30 & {\zsun} & $-$2.8$\pm$0.2 & 0.01$\pm$0.01 & 6.2$\pm$0.1 & 40$\pm$6 \\
NGC\,5194     &  3.5$\pm$1.0 & 2.35$^c$ & 2{\zsun} & 1$\pm$1 & 0.25$\pm$0.15 & 7$\pm$1 & 5$\pm$3 \\
NGC\,5253     &  5.5$\pm$1.0 & 2.35$^c$ & 0.4{\zsun} & $-$1.6$\pm$0.5 & 0.05$\pm$0.05 & 5.4$\pm$0.5 & 11$\pm$2 \\
NGC\,7496     &  2.5$\pm$0.5 & 2.35$^c$ & 2{\zsun} & $-$1.8$\pm$0.5 & 0.03$\pm$0.02 & $<$5.7$\pm$0.2 & 2.5$\pm$0.3 \\
NGC\,7673     &  6.5$\pm$0.5 & 2.35 & {\zsun} & $-$0.9$\pm$0.7 & 0.07$\pm$0.07 & 7$\pm$1 & 7.3$\pm$0.7 \\
NGC\,7714     &  4.5$\pm$0.3 & 2.35$^c$ & {\zsun} & $-$0.9$\pm$0.5 & 0.1$\pm$0.1 & 8$\pm$1 & 8.5$\pm$0.9 \\
Tol\,0440$-$381 &  5.0$\pm$1.0 & 2.35$^c$ & {\zsun} & 0$\pm$2 & 0.20$\pm$0.15 & 9$\pm$2 & 2.0$\pm$0.2$^e$ \\
\hline
\end{tabular}
\end{minipage}
\end{table*}

\begin{table*}
 \centering
 \begin{minipage}{140mm}
 \caption{Predicted Parameters for the Young Stellar Populations based on the FUV}
 \label{predictions}
 \begin{tabular}{@{}llllccl@{}}
 \hline
Galaxy & Log[\#O] & WR/O & Log[F(H$\alpha$)]\footnote{Fluxes in ergs s$^{1}$ cm$^{-2}$ \AA$^{-1}$.} & EW(H$\alpha$) & EW(4686) & Log[F(5500)]$^a$  \\
  &  &  &  & [\AA] & [\AA] & \\
\hline
IRAS\,08339$+$6517 & 0 & 0 & $-$10$^{+2}_{-2}$ & 22$^{+113}_{-8}$ & 0 & $-$11$^{+2}_{-2}$ \\
IRAS\,19245$-$4140 & 4$^{+2}_{-0}$ & 0.07$^{+0.01}_{-0.07}$ & $-$12$^{+2}_{-0}$ & 460$^{+600}_{-280}$ & 3$^{+14}_{-3}$ & $-$15$^{+2}_{-0}$ \\
IZw\,18       & 0 & 0 & $-$14$^{+1}_{-0}$ & 13$^{+95}_{-10}$ & 0 & $-$15$^{+1}_{-0}$ \\
M\,83         & 3.8$\pm$0.8 & 0.3$^{+0.3}_{-0.1}$ & $-$10.8$\pm$0.8 & 550$^{+370}_{-70}$ & 18$^{+15}_{-6}$ & $-$13.3$\pm$0.8 \\
MRK\,054      & 0 & 0 & $-$14$^{+2}_{-0}$ & 11$^{+6}_{-0}$ & 0 & $-$15$^{+2}_{-0}$ \\
MRK\,153      & 4.6$\pm$0.6 & 0 & $-$12.0$\pm$0.6 & 300$^{+100}_{-0}$ & 0 & $-$14.3$\pm$0.6 \\
NGC\,1140     & 4.6$\pm$0.8 & 0.1$^{+0.2}_{-0}$ & $-$11.5$\pm$0.8 & 200$^{+400}_{-0}$ & 15$^{+14}_{-9}$ & $-$13.7$\pm$0.8 \\
NGC\,1667     & $<$5.0$^{+0.5}_{-1}$ & 0.6$^{+2}_{-0.4}$ & $-$11.9$^{+0.5}_{-1}$ & 480$^{+440}_{-0}$ & 80$^{+0}_{-70}$ & $-$14.4$^{+0.5}_{-1}$ \\
NGC\,1672     & $<$5$\pm$1 & 0.3$^{+0.4}_{-0.1}$ & $-$11$\pm$1 & 550$^{+700}_{-70}$ & 20$^{+60}_{-10}$ & $-$14$\pm$1 \\
NGC\,1705     & 0 & 0 & $-$13.3$^{+0.2}_{-0}$ & 5$^{+15}_{-3}$ & 0 & $-$13.7$^{+0.2}_{-0}$ \\
NGC\,3310     & 0 & 0 & $-$13.2$\pm$0.5 & 0.5$^{+2}_{-0.3}$ & 0 & $-$13.1$\pm$0.5 \\
NGC\,3504     & 6.5$\pm$0.5 & 0.1$^{+0.2}_{-0.1}$ & $-$9.5$\pm$0.5 & 1250$^{+750}_{-700}$ & 7$^{+10}_{-7}$ & $-$12.3$\pm$0.5 \\
NGC\,3690     & 5.0$\pm$0.5 & 1$^{+0}_{-1}$ & $-$11.1$\pm$0.5 & 140$^{+180}_{-120}$ & 6$^{+1}_{-6}$ & $-$13.1$\pm$0.5  \\
NGC\,3991     & 8$\pm$1 & 0.07$^{+0.08}_{-0}$ & $-$9$\pm$1 & 500$^{+3000}_{-0}$ & 16$^{+28}_{-0}$ & $-$12$\pm$1 \\
NGC\,4194     & 7.0$\pm$1 & 0.3$^{+0.6}_{-0.2}$ & $-$9.0$\pm$1 & 520$^{+0}_{-270}$ & 11$^{+9}_{-4}$ & $-$12$\pm$1 \\
NGC\,4214     & 3.9$\pm$0.3 & 0.15$^{+0.15}_{-0.15}$ & $-$10.4$\pm$0.3 & 400$^{+500}_{-300}$ & 10$^{+15}_{-7}$ & $-$13.2$\pm$0.3 \\
NGC\,4449-Nuc & 0 &  0 &  $-$13.2$^{+0.2}_{-0}$ & 11$^{+6}_{-0}$ & 0 & $-$14.0$^{+0.1}_{-0}$ \\
NGC\,4449-HII & 2.7$^{+0.2}_{-0}$ & 0.12$^{+0.18}_{-0}$ & $-$11.8$^{+0.2}_{-0}$ & 247$^{+500}_{-0}$ & 20$^{+24}_{-9}$ & $-$13.9$^{+0.2}_{-0}$ \\
NGC\,5194     & 5$\pm$2 & 0.3$^{+0.4}_{-0.2}$ & $-$10$\pm$2 & 550$^{+700}_{-70}$ & 20$^{+30}_{-10}$ & $-$13$\pm$2 \\
NGC\,5253     & 3.0$\pm$0.5 & 0.15$^{+0.15}_{-0.15}$ & $-$11.5$\pm$0.5 & 400$^{+500}_{-300}$ & 10$^{+15}_{-7}$ & $-$14.2$\pm$0.5 \\
NGC\,7496     & $<$3.1$^{+0.5}_{-0}$ & 0.10$^{+0.10}_{-0.08}$ &  $-$12.6$^{+0.5}_{-0}$ & 1250$^{+380}_{-330}$ & 7$\pm$5 & $-$15.5$^{+0.5}_{-0}$ \\
NGC\,7673     & 3$^{+2}_{-0}$ &  1$^{+0}_{-1}$ & $-$13$^{+2}_{-0}$ & 140$^{+180}_{-120}$ & 6$^{+1}_{-6}$ & $-$15$^{+2}_{-0}$ \\
NGC\,7714     & 5$\pm$1 & 0.19$^{+0}_{-0.07}$ & $-$11$\pm$1 & 748$^{+0}_{-500}$ & 44$^{+0}_{-28}$ & $-$14$\pm$1 \\
TOL\,0440-381 & 6$^{+2}_{-1}$ & 0.1$^{+0.8}_{-0.1}$ & $-$12$^{+2}_{-1}$ & 250$^{+500}_{-0}$ & 20$^{+24}_{-13}$ & $-$14$^{+2}_{-1}$ \\
\hline
\end{tabular}
\end{minipage}
\end{table*}

\begin{figure*}
\includegraphics[width=14cm]{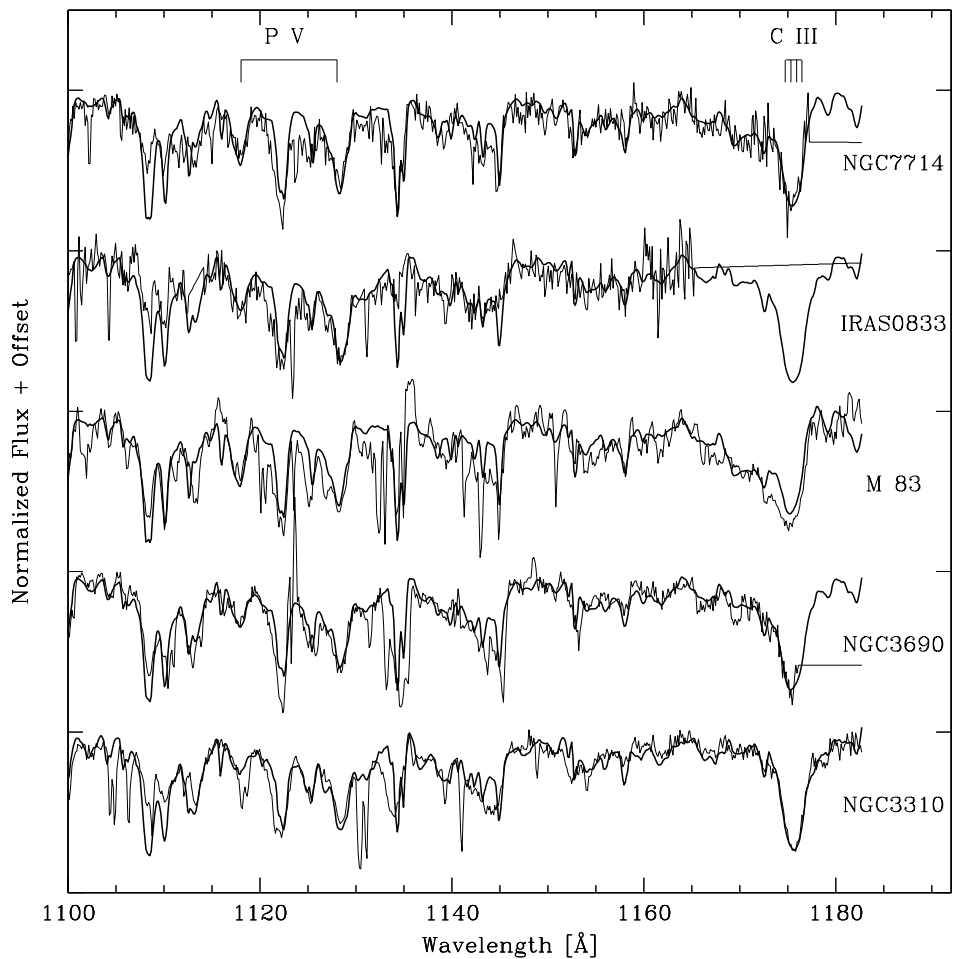}
\caption{ FUSE spectra of well-known starburst galaxies. 
Thin lines represent the FUSE spectra and thick lines
are the best {\tt LavalSB} synthetic spectra obtained for an
instantaneous bursts with a standard IMF (i.e. $\alpha=2.35$ and
stars from 1 to 100\,{\msun}).
From top to bottom:
NGC\,7714 superimposed on a model having 4.5\,Myr and {\zsun};
IRAS\,08339+6517 and a model at 7.0\,Myr and {\zsun};
M\,83 and a model at 3.5\,Myr and 2\,{\zsun};
NGC\,3690 and a model at 6.5\,Myr and {\zsun};
and NGC\,3310 superimposed on the best-fitting model having 
18\,Myr at 2\,{\zsun} and $\alpha$(IMF)=2.35. }
\label{figproto}
\end{figure*}

\begin{figure*}
\includegraphics[width=16cm]{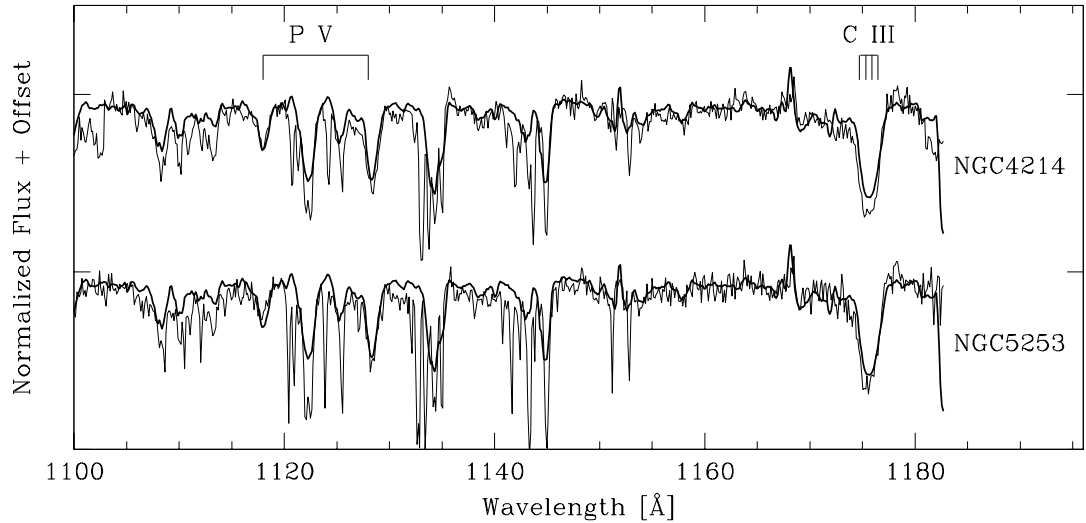}
\caption{FUSE spectra of NGC\,4214 and NGC\,5253.
Thin lines represent the FUSE spectra and thick lines
are {\tt LavalSB} synthetic spectra obtained for a single instantaneous burst with
a standard IMF (i.e. $\alpha=2.35$ and stars from 1 to 100\,{\msun}).
Top: NGC\,4214 superimposed on a model having 4.5\,Myr and 0.4\,{\zsun}. 
Bottom: NGC\,5253 superimposed on a model having 4.5\,Myr and 0.4\,{\zsun}.}
\label{figmetal}
\end{figure*}

\begin{figure*}
\includegraphics[width=12cm]{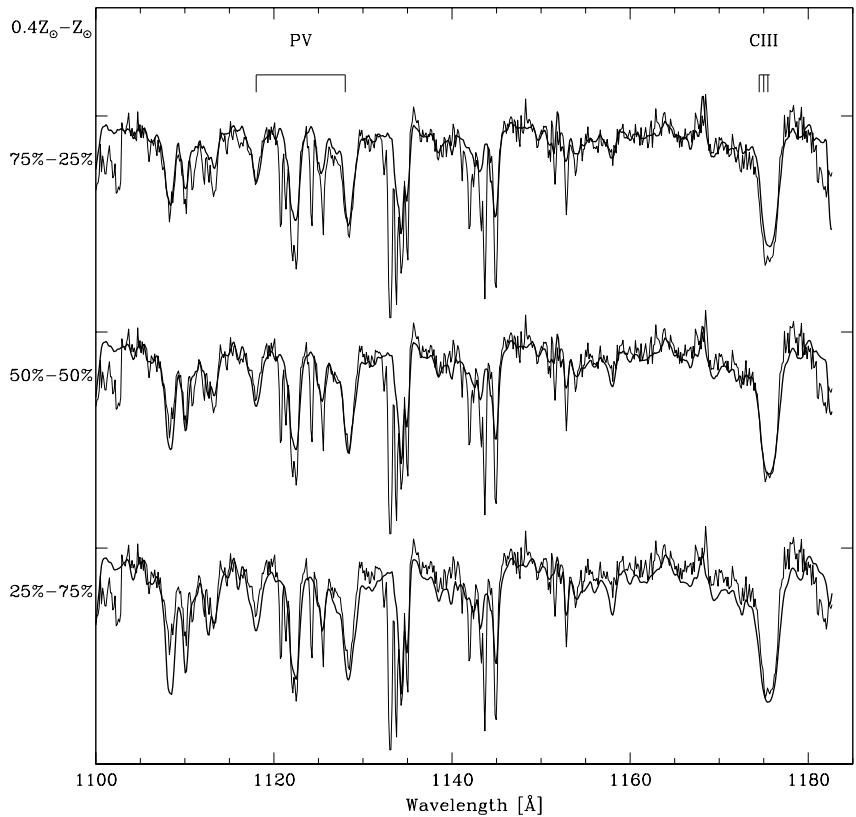}
 \caption{FUSE spectrum of NGC\,4214 superimposed on two burst models combining 0.4\,{\zsun} and {\zsun} metallicities. Both synthetic populations have 5.5\,Myr and a standard IMF. Flux proportions for each combination are indicated on the left, in percents. Thick line: result of combinations. Thin line: NGC\,4214 spectrum.}
\label{fig4214}
\end{figure*}

\begin{figure*}
\includegraphics[width=12cm]{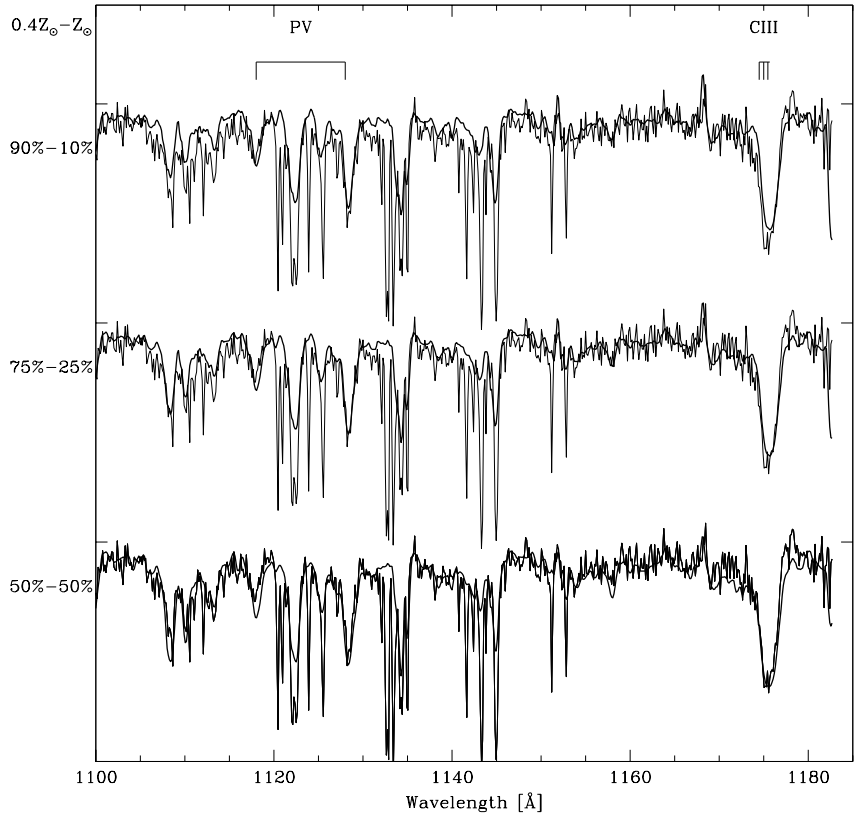}
\caption{FUSE spectrum of NGC\,5253 superimposed on two burst models combining 0.4\,{\zsun} and {\zsun} metallicities. Both synthetic populations have 5.5\,Myr and a standard IMF. See Figure~\ref{fig4214} for legend.}
\label{fig5253}
\end{figure*}

\begin{figure*}
\includegraphics[width=16cm]{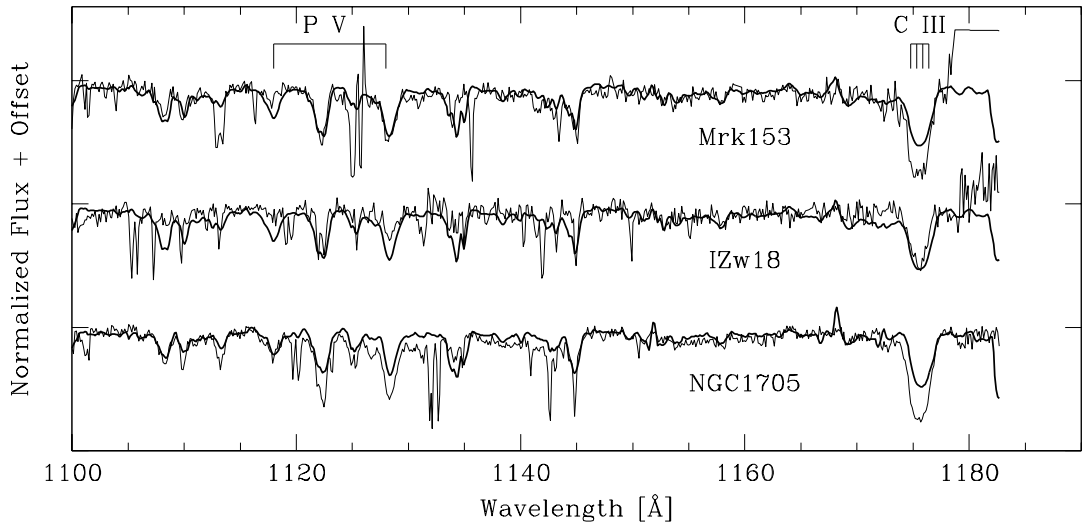}
\caption{FUSE spectra of starbursts with low metallicity. From top to bottom: 
Mrk\,153 superimposed on a model having 6.5\,Myr and 0.2\,{\zsun};
NGC\,1705 superimposed on a model having 7.0\,Myr at 0.4\,{\zsun} and $\alpha$(IMF)=2.35;
IZw\,18 superimposed on a model having 7.0\,Myr at 0.2\,{\zsun} and $\alpha$(IMF)=2.2.
Thin lines: FUSE spectra; thick lines: {\tt LavalSB} synthetic spectra of instantaneous bursts.}
\label{figlowz}
\end{figure*}

\begin{figure*}
\includegraphics[width=16cm]{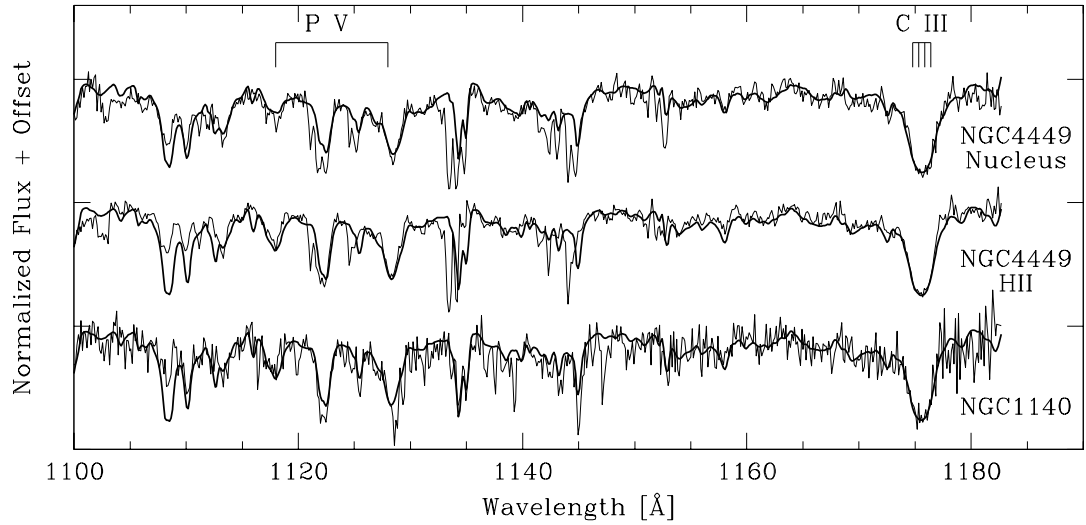}
\caption{FUSE spectra of starbursts with steep IMF. From top to bottom: 
Nucleus of NGC\,4449 superimposed on the best-fitting model having 10.0\,Myr at {\zsun} and $\alpha$(IMF)=3.3;
{\hii} region of NGC\,4449 superimposed on the best-fitting model having 5.0\,Myr at {\zsun} and $\alpha$(IMF)=3.3;
NGC\,1140 superimposed on the best-fitting model having 5.0\,Myr at {\zsun} and $\alpha$(IMF)=2.8.
Thin lines: FUSE spectra; thick lines: {\tt LavalSB} synthetic spectra of instantaneous bursts.}
\label{imffig}
\end{figure*}

\begin{figure*}
\includegraphics[width=18.2cm]{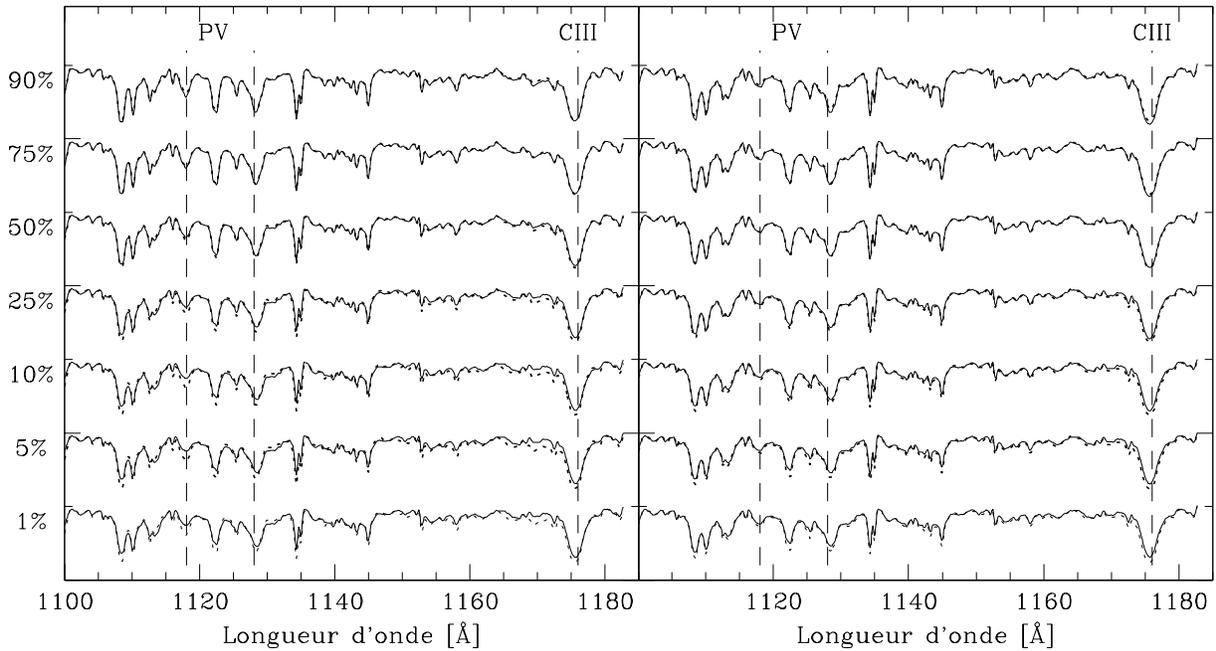}
\caption{Dilution of a 5.0\,Myr burst (left panel) and of a 10.0\,Myr burst (right panel) with a 20\,Myr burst. Solid line: combination of two single population models of 5.0 and 20\,Myr with both standard IMF. The FUV flux proportion of the youngest population is indicated on the left. Dotted line: single synthetic population of 5.0 and 10.0\,Myr having $\alpha$(IMF)=3.3.}
\label{figdilution}
\end{figure*}

\begin{figure*}
\includegraphics[width=18cm]{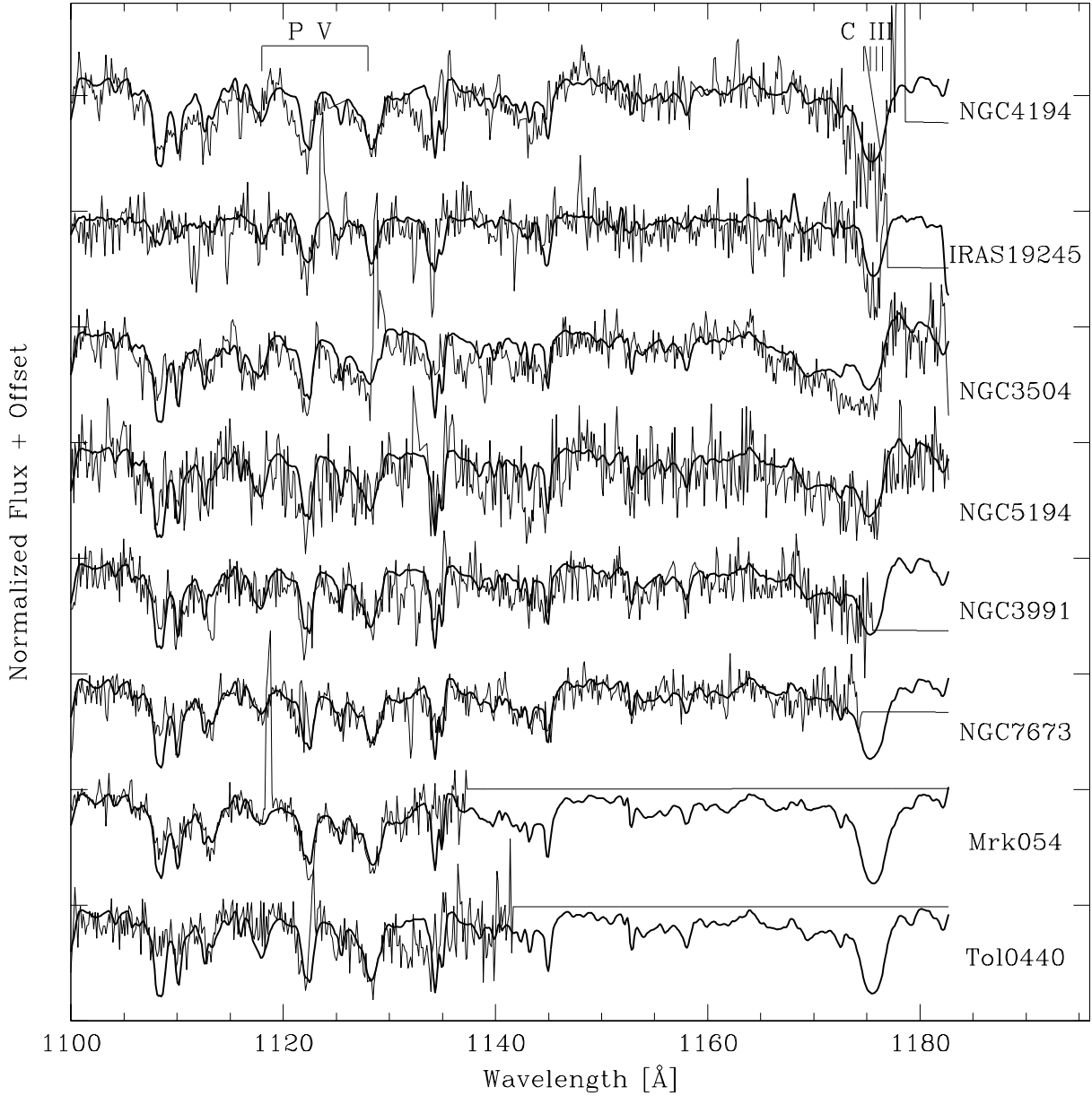}
\caption{FUSE spectra of starbursts with lower signal-to-noise ratios. From top to bottom:
NGC\,4194 superimposed on the best-fitting model having 5.5\,Myr at {\zsun} and $\alpha$(IMF)=2.35;
IRAS\,19245-4140 superimposed on the best-fitting model having 5.0\,Myr at 0.4\,{\zsun} and $\alpha$(IMF)=2.35;
NGC\,3504 superimposed on the best-fitting model having 2.5\,Myr at 2\,{\zsun} and $\alpha$(IMF)=2.35;
NGC\,5194 (nucleus$+${\hii} region) superimposed on the best-fitting model having 3.5\,Myr at 2\,{\zsun} and $\alpha$(IMF)=2.35;
NGC\,3991 superimposed on the best-fitting model having 4.0\,Myr at {\zsun} and $\alpha$(IMF)=2.8;
NGC\,7673 superimposed on the best-fitting model having 6.5\,Myr at {\zsun} and $\alpha$(IMF)=2.35
Mrk\,054 superimposed on the best-fitting model having 10.0\,Myr at {\zsun} and $\alpha$(IMF)=2.35;
and Tol\,0440-381 superimposed on the best-fitting model having 5.0\,Myr at {\zsun} and $\alpha$(IMF)=2.35.
Thin line: FUSE spectra; thick line {\tt LavalSB} synthetic spectra of instantaneous bursts.}
\label{figsn}
\end{figure*}

\begin{figure*}
\includegraphics[width=14cm]{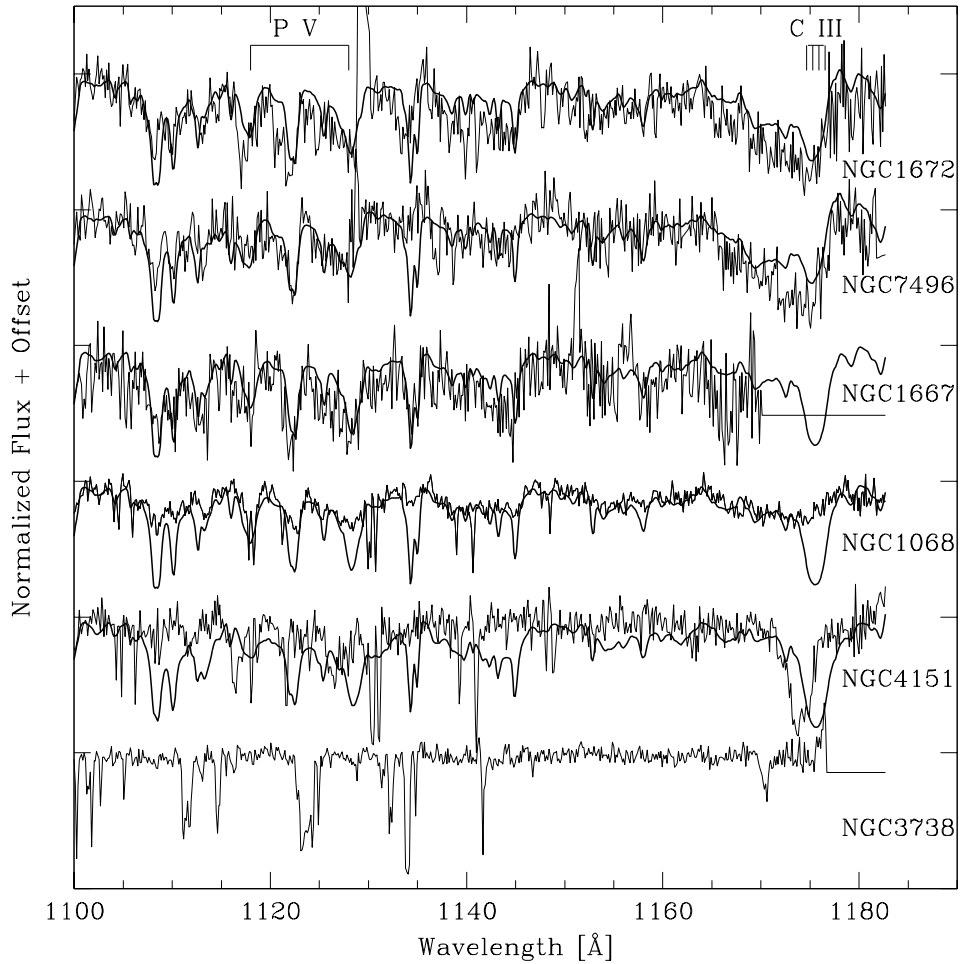}
\caption{FUSE spectra of Seyfert galaxies. From top to bottom:
NGC\,1672 superimposed on the best-fitting model having 3.5\,Myr at 2\,{\zsun} and $\alpha$(IMF)=2.35; 
NGC\,7496 superimposed on the best-fitting model having 2.5\,Myr at 2\,{\zsun} and $\alpha$(IMF)=2.35;
NGC\,1667 superimposed on the best-fitting model having 4.0\,Myr at 2\,{\zsun} and $\alpha$(IMF)=2.35;
NGC\,1068 superimposed with a {\zsun} model for comparison;
NGC\,4151 in a low-level activity phase with a {\zsun} model for comparison;
NGC\,3783 with no stellar line. In NGC\,4151, the massive stellar content (if any) is diluted by the non-thermal radiation (see text).
Thin line: FUSE spectra; thick line {\tt LavalSB} synthetic spectra of instantaneous bursts.}
\label{figagn}
\end{figure*}

\label{lastpage}

\end{document}